\documentclass[%
 reprint,
nofootinbib,
 amsmath,amssymb,
 aps,
showkeys
]{revtex4-1}

\usepackage{tgtermes}

\usepackage{graphicx}
\usepackage{dcolumn}
\usepackage{bm}
\usepackage{hyperref}
\usepackage{braket}
\usepackage[normalem]{ulem}

\DeclareMathOperator{\sgn}{sgn}
\newcommand{\D}{\mathrm{d}}				

\usepackage{xcolor}						

\begin{document}

\preprint{APS/123-QED}

\title{Black holes, stationary clouds and magnetic fields}%

\author{Nuno M. Santos$^{1,2,}$}
\email{herdeiro@ua.pt}
\author{and Carlos A. R. Herdeiro$^{2,}$}
\email{nunomoreirasantos@tecnico.ulisboa.pt}
\affiliation{$^1$Centro de Astrof\'{\i}sica e Gravita\c c\~ao  --- CENTRA,
Departamento de F\'{\i}sica, Instituto Superior T\'ecnico --- IST,
Universidade de Lisboa --- UL,
Av. Rovisco Pais 1, 1049-001 Lisboa, Portugal}
\affiliation{$^2$Centre for Research and Development  in Mathematics and Applications (CIDMA) and \\
Departamento de Matem\'atica da Universidade de Aveiro \\
Campus de Santiago, 3810-183 Aveiro, Portugal
}%

\date{January 2021}

\begin{abstract}
As the electron in the hydrogen atom, a bosonic field can bind itself to a black hole occupying a discrete infinite set of states. When (i) the spacetime is prone to superradiance and (ii) a confinement mechanism is present, some of such states are infinitely long--lived. These equilibrium configurations, known as stationary clouds, are states ``synchronized" with a rotating black hole's event horizon. For most, if not all, stationary clouds studied in the literature so far, the requirements (i)--(ii) are independent of each other. However, this is not always the case. This paper shows that massless neutral scalar fields can form stationary clouds around a Reissner--Nordstr\"{o}m black hole when both are subject to a uniform magnetic field. The latter simultaneously enacts both requirements by creating an ergoregion (thereby opening up the possibility of superradiance) and trapping the scalar field in the black hole's vicinity. This leads to some novel features, in particular, that only black holes with a  subset of the possible charge to mass ratios can support stationary clouds.
\end{abstract}

\maketitle


\section{\label{sec:level1}Introduction}

Neutron stars and black holes in binary systems feed some of the most powerful astrophysical events in the Universe. Their gravitational--wave luminosity can reach a peak of approximately $10^{57}$ erg s$^{-1}$~\cite{LIGOScientific:2018mvr,Cardoso:2018nkg}, only comparable to the electromagnetic luminosity of the most luminous gamma--ray bursts~\cite{Frederiks:2013cga}. The Advanced LIGO/Virgo's first and second observation runs reported the detection of gravitational waves from ten different binary black hole mergers and a single binary neutron star merger. During the first half of the third observing run, a total of 39 gravitational--wave candidate events were observed, three of which may have originated from neutron star--black hole mergers~\cite{Abbott:2020niy}. Joint detections of gravitational and electromagnetic waves from neutron star--black hole coalescences are of particular interest for constraining the equation of state of dense nuclear matter~\cite{Raaijmakers:2019dks} and measuring the Hubble constant~\cite{Vitale:2018wlg}. Furthermore, some neutron stars, known as magnetars, are endowed with super--strong magnetic fields reaching $10^{12}$--$10^{15}$ G~\cite{Olausen:2013bpa}. For instance, the magnetar SGR J1745--2900, which orbits the supermassive black hole Sagittarius A$^*$, has a surface dipolar magnetic field of $10^{14}$ G. Neutron star--black hole binary systems are thus natural laboratories for probing the intricate interaction of black holes with magnetic fields. 

A magnetic field $B_0$ permeating a black hole with mass $M$ curves the spacetime in a non-negligible way beyond a threshold value set by $MB\sim 1$~\cite{Frolov:2010mi}, or reinstating familiar units
\begin{align}
B\equiv\frac{c^4}{G^{3/2}M}\sim 10^{19}\left(\frac{M_\odot}{M}\right)~\text{G}\ ,
\label{eq:1.1}
\end{align}
where $M_\odot$ is the solar mass. A magnetic field of order $B$ or larger warps significantly spacetime in the vicinity of the event horizon (without changing its topology). Since the field strength of a magnetic dipole falls off as the cube of the distance from it, it is unlikely that stellar--mass black holes or even supermassive black holes are subject to magnetic fields of order $B$. 
 
Even if its strength is significantly smaller than $B$, the impact of a magnetic dipole on fields interacting with black holes may be non--negligible, as they can acquire an effective mass and be trapped in its vicinity. A massless field traversing the black hole's vicinity would then behave as if it had non--vanishing mass and its effective mass would depend on the magnetic field strength. In addition, if the field is bosonic, it can induce black--hole superradiance, i.e. the extraction of energy and angular momentum from rotating black holes (for a review, see~\cite{Brito:2015oca}). Black--hole superradiance takes place when the phase angular velocity~$w$ of the bosonic field satisfies
\begin{align}
w<m\Omega_\mathcal{H}\ ,
\label{eq:1.2}
\end{align}
where~$m$ is the azimuthal harmonic index and $\Omega_\mathcal{H}$ is the black hole's angular velocity. Together with a natural confinement mechanism, black--hole superradiance is responsible for bosonic fields to form quasi--bound states. These are constinuously fed the extracted black hole's energy and angular momentum until Eq.~\eqref{eq:1.2} saturates, i.e. $w=m\Omega_\mathcal{H}$, and they become bound states. The new equilibrium state is expected to be a classical bosonic condensate in equilibrium with the slowed--down black hole, which for a complex bosonic field is a hairy black hole~\cite{Herdeiro:2014goa,Herdeiro:2016tmi,East:2013mfa,Herdeiro:2017phl,Santos:2020pmh}. 

The bosonic field remains trapped in the vicinity of the black hole when it is massive. A non--vanishing intrinsic mass, however, is not always mandatory. Trapping can be attained even when the field is massless. For instance, a massless bosonic field interacting with a black hole immersed in a magnetic field is likely to form bound states. The magnetic field creates a potential barrier, confining the field into the neighborhood of the black hole.


An example that naturally embodies this idea is the interaction of a massless scalar field with a Reissner--Nordstr\"{o}m black hole embedded in a uniform axial magnetic field~\footnote{Although this is not a realistic astrophysical scenario, it suffices to sketch the main argument of the paper.}. The latter is described by the Reissner--Nordstr\"{o}m--Melvin (RNM) solution~\cite{Ernst:1976mzr,Gibbons:2013yq}, obtained via a solution--generating technique known as Harrison (or ``magnetizing") transformation. Interestingly, the RNM solution is a stationary (rather than a static) solution of the Einstein--Maxwell theory . The rotation is sourced by the coupling between the black hole's electric charge and the external magnetic field. Besides, the spacetime features an ergoregion and, as a result, is prone to black--hole superradiance even for electrically neutral bosonic fields. This contrasts with the case of asymptotically--flat Reissner--Nordstr\"{o}m black holes wherein (charged) superradiance is possible but only for \textit{charged} bosonic fields~\cite{Bekenstein:1973mi} and a superradiant instability does not follow from a mass term; it requires, for instance, enclosing the black hole with a reflecting mirror -- see, \textit{e.g.},~\cite{Herdeiro:2013pia,Degollado:2013bha,Sanchis-Gual:2015lje}.

The present paper focuses on bound states between a massless scalar field and a RNM black hole (cf.~\cite{Vieira:2016ubt}). These real--frequency states are characterized by the threshold of superradiance $w=m\Omega_\mathcal{H}$, hereafter referred to as \textit{synchronisation condition}, and were first reported in~\cite{Hod:2012px}, in which the author named them~\textit{stationary clouds}. Much attention has been paid to such synchronized states since their discovery \cite{Hod:2013zza,Hod:2014baa,Benone:2014ssa,Wang:2015fgp,Hod:2015goa,Siahaan:2015xna,Hod:2016lgi,Hod:2016yxg,Huang:2016qnk,Bernard:2016wqo,Sakalli:2016xoa,
Ferreira:2017cta,Richartz:2017qep,Huang:2017whw,Huang:2018qdl,Garcia:2018sjh,Delgado:2019prc,Kunz:2019bhm,Garcia:2019zla,Santos:2020pmh,Santos:2020sut}, yet most works rely on intrinsically massive {fields}. For the case under consideration here, the fields need not have a non-vanishing mass for stationary clouds to arise.\footnote{The same is true for {AdS} asymptotics -- see, \textit{e.g.},~\cite{Wang:2015fgp}.} A peculiar feature of this model is that the scalar field's effective mass is proportional to the black hole's angular velocity, the proportionality constant being a function of the specific electric charge $Q/M$ alone, where $M$ and $Q$ are, respectively, the black hole's mass and electric charge. Curiously enough, the condition for the existence of bound states is only met for values of $Q/M$ in a subset of~$[-1,1]$. 

The paper is organized as follows. First, the Einstein--Maxwell theory minimally coupled to a complex, ungauged scalar field is introduced in~\autoref{sec:level2}. Together with a constant scalar field, the RNM solution is a particular case of the theory. Its main features are outlined in~\autoref{sec:level2.1}, followed by a linear analysis of scalar field perturbations in~\autoref{sec:level2.2}. The main results on stationary clouds are presented in~\autoref{sec:level3}. A summary of the work can be found in~\autoref{sec:level4}.

Natural units ($G=c=1$) are consistently used throughout the text. Additionally, the metric signature $(-,+,+,+)$ is adopted.

\section{\label{sec:level2}Framework}

The action for the Einstein--Maxwell theory minimally coupled to a complex\footnote{Stationary clouds are not exclusive to complex scalar fields. A single real scalar field can equally form infinitely long--lived states at linear level -- see~\cite{Herdeiro:2015gia}.}, ungauged scalar field $\Psi$ is
\begin{align}
\mathcal{S}=\frac{1}{4\pi}\int\D^4x\sqrt{-g}\left[\frac{R}{4}-\frac{F^2}{4}-(\nabla^\mu\Psi^*)(\nabla_\mu\Psi)\right]\ ,
\label{eq:2.1}
\end{align}
where $\bm{F}=\D\bm{A}$ is the electromagnetic tensor and~$\bm{A}$ is electromagnetic four--potential. 

The corresponding equations of motion read
\begin{align}
G_{\mu\nu}=2\left[T_{\mu\nu}^{(\bm{A})}+T_{\mu\nu}^{(\Psi)}\right]
\ ,\
\Box\Psi=0
\ ,\ 
\nabla_\mu F^{\mu\nu}=0\ ,
\end{align}
where $\Box\equiv\nabla_\mu\nabla^\mu$ is the d'Alembert operator and
\begin{align}
T_{\mu\nu}^{(\bm{A})}&\equiv{F_\mu}^\sigma F_{\nu\sigma}-\frac{1}{4}g_{\mu\nu}F_{\sigma\lambda}F^{\sigma\lambda} ,\\
T_{\mu\nu}^{(\Psi)}&\equiv2\partial_{(\mu}\Psi^*\partial_{\nu)}\Psi-g_{\mu\nu}(\partial_\lambda\Psi^*)(\partial^\lambda\Psi)\ 
\end{align}
are the stress--energy tensors of the electromagnetic and scalar fields, respectively. The action has a global $U(1)$ invariance with respect to the scalar field thanks to its complex character.

This field theory admits all of the stationary solutions of general relativity. These are characterized by $\Psi=\Psi_0$, for some constant~$\Psi_0$. Linearizing the equations of motion around $\Psi=\Psi_0$, one obtains the ordinary Einstein--Maxwell equations together with the Klein--Gordon equation for the scalar field perturbation $\delta\Psi\equiv(\Psi-\Psi_0)$. This system describes the linear or zero--backreaction limit of the theory: the limit in which the backreaction of both the gravitational and electromagnetic fields to a non--constant scalar field is negligible. This first--order approximation suffices to capture potentially relevant astrophysical phenomena such as superradiant scattering. The framework allows one to solve the Klein--Gordon equation $\Box(\delta\Psi)=0$ for a known solution $\{\bm{g},\bm{A}\}$ of the Einstein--Maxwell equations.

\subsection{\label{sec:level2.1}Reissner--Nordstr\"{o}m--Melvin black holes}

This paper will focus on scalar field perturbations of RNM black holes. These solutions belong to a family of electrovacuum type D solutions of the Einstein--Maxwell equations which asymptotically resemble the magnetic Melvin universe. The latter describes a non--singular, static, cylindrically symmetric spacetime representing a bundle of magnetic flux lines in gravitational--magnetostatic equilibrium. It can be loosely interpreted as Minkowski spacetime immersed in a uniform magnetic field; but it should be kept in mind that such magnetic field, no matter how small, changes the global structure of  the spacetime, in particular its asymptotics.

Given an asymptotically--flat, stationary, axi--symmetric solution of Einstein--Maxwell equations, it is possible to embed it in a uniform magnetic field via a solution--generating technique called Harrison transformation (also commonly known as ``magnetizing''  transformation). This possibility, first realized by Harrison~{\cite{Harrison:1971}}, was explored for the Schwarzschild and Reissner-Nordstr\"{o}m solutions~\cite{Ernst:1976mzr} and for the Kerr and Kerr--Newman solutions~\cite{Ernst:1976a}. 

The RNM solution, which describes a Reissner--Nordstr\"{o}m black hole permeated by a uniform magnetic field, reads~\cite{Gibbons:2013yq}
\begin{align}
&\begin{aligned}
\bm{g}=|\Lambda|^2\left(-\frac{\Delta}{r^2}\right.\left.\bm{\D}t^2+\frac{r^2}{\Delta}\bm{\D}r^2+r^2\bm{\D}\theta^2\right)\nonumber\\
+\frac{r^2\sin^2\vartheta}{|\Lambda|^2}(\bm{\D}\varphi-\Omega\bm{\D}t)^2\ ,\\
\end{aligned}\\
&\bm{A}=\Phi_0\bm{\D}t+\Phi_3(\bm{\D}\varphi-\Omega\bm{\D}t)
\end{align}
where $t\in(-\infty,+\infty)$, $r\in(0,+\infty)$, $\vartheta\in[0,\pi]$, $\varphi\in[0,2\pi)$ and
\begin{align*}
&\Delta=r^2-2M r+Q^2\ ,\\
&\Lambda=1+\frac{1}{4}B_0^2(r^2\sin^2\vartheta+Q_0^2\cos^2\vartheta)-iQB_0\cos\vartheta\ ,\\
&\Omega=-\frac{2QB_0}{r}+\frac{QB_0^3 r}{2}\left(1+\frac{\Delta}{r^2}\cos^2\vartheta\right)\ ,\\
&\Phi_0=-\frac{Q}{r}+\frac{3}{4}QB_0^2r\left(1+\frac{\Delta}{r^2}\cos^2\vartheta\right)\ ,\\
&\Phi_3=\frac{2}{B_0}-\frac{1}{|\Lambda|^2}\left[\frac{2}{B_0}+\frac{B_0}{2}\left(r^2\sin^2\vartheta+3Q^2\cos^2\vartheta\right)\right]\ .
\end{align*}
$B_0$ is the strength of the magnetic field, which is assumed to be much weaker than the threshold value~\eqref{eq:1.1}, i.e $MB_0\ll MB=1$.

When applied to the Reissner--Nordstr\"{o}m solution, the Harrison transformation produces a stationary (rather than a static) solution. The dragging potential $\Omega$ is directly proportional to the coupling~$QB_0$, which suggests that the interaction between the charge $Q$ and the magnetic field $B_0$ serves as a source for rotation. 

The solution possesses two (commuting) Killing vectors, $\bm{\xi}=\bm{\partial}_t$ and $\bm{\eta}=\bm{\partial}_\varphi$, associated to stationarity and axi--symmetry, respectively. The line element has coordinate singularities at $\Delta=0$ when $Q^2\leq M^2$, which solves for $r_\pm=M\pm\sqrt{M^2-Q^2}$. The hypersurface $r=r_+$ ($r=r_-$) is the outer (inner) horizon. Besides, there is an ergo--region that extends to infinity along the axial direction, but not in the radial direction. Here, ergo--region means the regions outside the outer horizon wherein  $\bm{\xi}$ is spacelike.

The dragging potential $\Omega$ is constant (i.e. $\vartheta$--independent) on $r=r_+$, where it has the value
\begin{align}
\Omega_\mathcal{H}\equiv-\frac{2QB_0}{r_+}\left(1-\frac{r_+^2B_0^2}{4}\right)\ .
\label{eq:2.3}
\end{align}
$\Omega_\mathcal{H}$ is the angular velocity of the outer horizon. The Killing vector $\bm{\chi}=\bm{\xi}+\Omega_\mathcal{H}\bm{\eta}$ becomes null on the hypersurface $r=r_+$ and it is timelike outside it.

\subsection{\label{sec:level2.2}Scalar field perturbations}
In general, the Klein--Gordon equation $\Box(\delta\Psi)=0$ does not admit a multiplicative separation of variables of the form
\begin{align}
\delta\Psi(t,\bm{r})=e^{-iwt}R(r)S(\vartheta)e^{+im\varphi}\ ,
\label{eq:3.2}
\end{align}
where $w$ is the phase angular velocity, $R$ and $S$ are respectively the radial and angular fucntions and $m\in\mathbb{Z}$ is the azimuthal harmonic index. However, in the limit of sufficiently ``weak'' magnetic fields, i.e. neglecting terms of order\footnote{For a straightforward identification of the order of each term, it is convenient to introduce the dimensionless quantities~$\{tB_0,rB_0,MB_0,QB_0,w/B_0\}$ so that all physical quantities are measured in units of the magnetic field strength. Note that the first four quantities are of order~$\mathcal{O}(B_0)$,  whereas the last is of order $\mathcal{O}(B_0^{-1})$.} higher than $\mathcal{O}(B_0^2)$ , the ansatz~\eqref{eq:3.2} actually reduces the problem to two differential equations in the coordinates $r$ and $\vartheta$. The radial and angular equations read~\cite{Vieira:2016ubt}
\begin{align}
&\frac{\D}{\D r}\left(\Delta\frac{\D R}{\D r}\right)+\left[\frac{K^2}{\Delta}-(m^2B_0^2r^2+\lambda)\right]R=0\ ,
\label{eq:3.3a}
\\
&\frac{1}{\sin\vartheta}\frac{\D}{\D\vartheta}\left(\sin\vartheta\frac{\D S}{\D\vartheta}\right)\nonumber\\
&\quad~\quad~+\left(\lambda-\frac{m^2}{\sin^2\vartheta}-3m^2Q^2B_0^2\cot^2\vartheta\right)S=0\ ,
\label{eq:3.3b}
\end{align}
respectively, where $K=r^2w+2mQB_0r$ and $\lambda$ is the separation constant. Equations~\eqref{eq:3.3a}--\eqref{eq:3.3b} are both confluent Heun equations: the former (latter) has singular points at $r=r_\pm$ ($\vartheta=0,\pi$). They are coupled via the Killing eigenvalues $\{w,m\}$, $B_0$, $Q$ and the separation constant $\lambda$ and remain invariant under the discrete transformation $\{w,mQB_0\}\rightarrow\{-w,-mQB_0\}$. This guarantees that, without loss of generality, one can take $\sgn(w)=\sgn(B_0)=+1$. When $mQB_0=0$, the angular equation reduces to the general Legendre equation, whose canonical solutions are the associated Legendre polynomials of degree $\ell$ and order $m$, $P_{\ell}^m(\vartheta)$, provided that $\lambda=\ell(\ell+1)$. Thus, if $|mQB_0|\ll1$, the angular dependence of $\delta\Psi$ is approximately described by the scalar spherical harmonics of degree $\ell$ and order $m$, $Y_\ell^m(\vartheta,\varphi)=P_\ell^m(\vartheta)e^{+im\varphi}$.

Equation~\eqref{eq:3.3a} can be cast in Schr\"{o}dinger--like form, yielding
\begin{align}
-\frac{\D^2\rho}{\D y^2}+V_\text{eff}(y)\rho=w^2\rho\ ,
\end{align}
where $\rho\equiv rR$ and $y$ is the tortoise coordinate, defined by
\begin{align*}
y(r)=r+\frac{r_+^2}{r_+-r_-}\log(r-r_+)-\frac{r_-^2}{r_+-r_-}\log(r-r_-)\ ,
\end{align*}
which maps the interval $r\in[r_+,\infty)$ into $r_*\in(-\infty,+\infty)$. The effective potential~$V_\text{eff}$, whose expression is omitted here, has the following limiting behavior:
\begin{align}
&\lim_{y\rightarrow-\infty}V_\text{eff}(y)={w^2-(w-m\Omega_\mathcal{H})^2}\ ,\\
\quad
&\lim_{y\rightarrow+\infty}V_\text{eff}(y)=m^2B^2\ .
\end{align}

The last limit suggests that a non--vanishing external magnetic field makes the scalar field acquire an effective mass $\mu_\text{eff}=\sqrt{m^2B_0^2}$. It is important to remark, however, that the problem at hand is not equivalent to that of a massive scalar field perturbation on an asymptotically--flat stationary spacetime, wherein the mass dominates the asymptotic behavior of the field. Besides providing the field an effective mass, the magnetic field also changes the asymptotic behavior at infinity (to be that of the Melvin magnetic universe), which has similarities with AdS asymptotics in the sense that it is naturally confining.

\begin{figure}[h]
\includegraphics[width=\columnwidth]{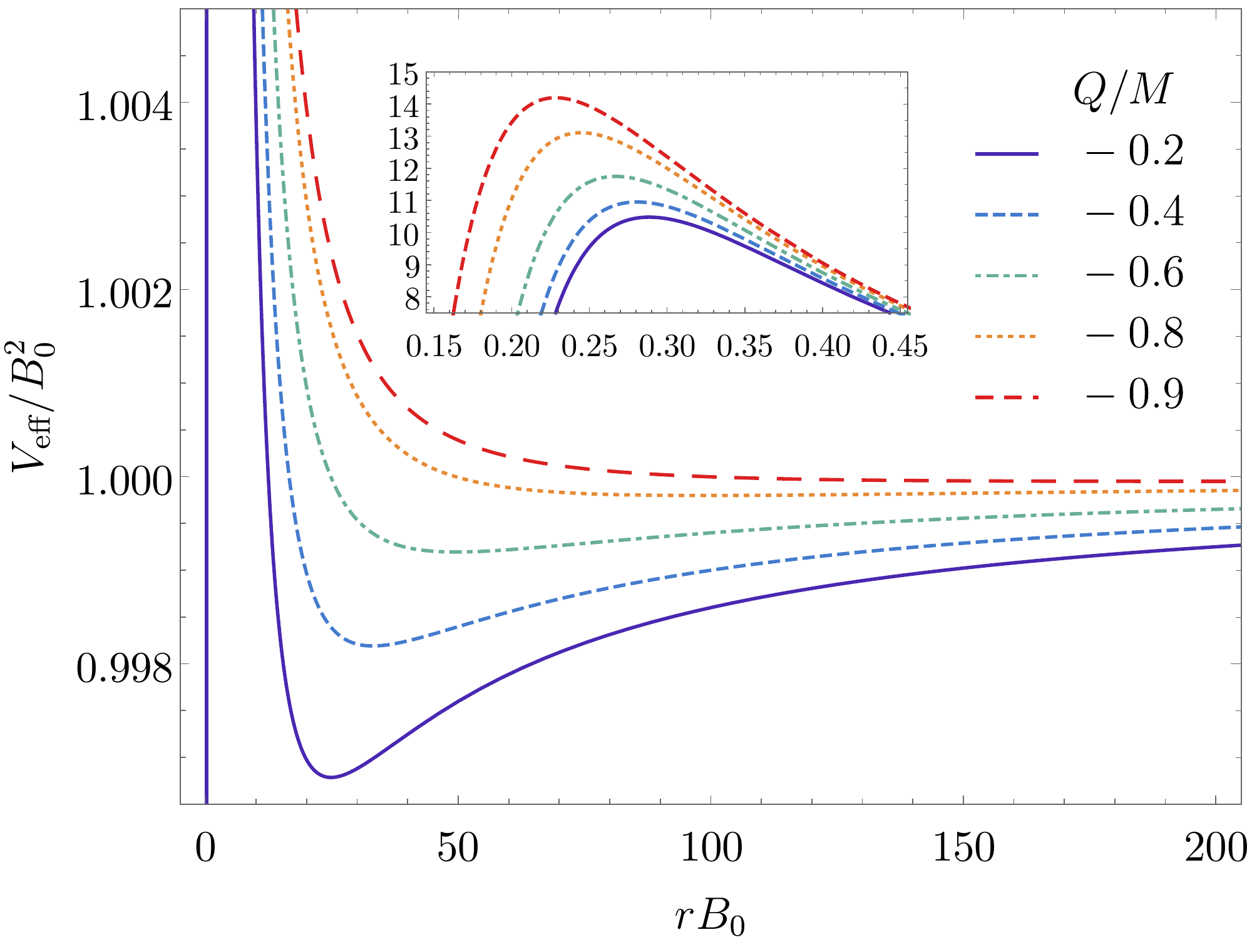}
\caption{Effective potential for scalar field perturbations with $\ell=m=1$ and $w=0.5B_0$ of RNM black holes with $MB_0=0.1$. (Inset) Zoom near $rB_0\sim 1$ to display the maximum of the effective potential.}
\label{fig:0}
\end{figure}

\autoref{fig:0} shows the effective potential as a function of the radial coordinate $r$ for different (negative) specific electric charges. In an asymptotically--Melvin spacetime, the magnetic field acts like a potential barrier at $rB_0\sim1$, whose maximum, about ten times larger than $\mathcal{O}(B_0^2)$, approaches the outer horizon with decreasing $Q/M$ (i.e. tending to extremality). Moreover, there is a potential well for all positive specific electric charges (not plotted in~\autoref{fig:0}) as well as for negative ones above a certain threshold (away from extremality). The effective potential resembles a mirror placed at $rB_0\sim1$ and confines (low--frequency) scalar field perturbations in the black hole's vicinity~\cite{Konoplya:2008hj,Brito:2014nja}. It is then natural to impose a Robin (or mixed) boundary condition at $r=r_0$ as the outer boundary condition, 
\begin{align}
\tan(\zeta)=-\frac{R(r_0)}{R'(r_0)}\ ,
\label{eq:4.3}
\end{align}
where $r_0$ is of order~$\mathcal{O}(B_0^{-1})$, $\zeta\in[0,\pi)$, with $\zeta=0$ ($\zeta=\pi/2$) corresponding to a Dirichlet (Neumann) boundary condition, and the prime denoting differentiation with respect to $r$.

In realistic astrophysical scenarios, magnetic fields occur in accretion disks around black holes. The ``magnetic" potential barrier is then at a radial distance smaller than about the mean radius $D$ of the disk, i.e. $r_0\lesssim D$. Since the matter in the accretion disk is expected to be close to the innermost stable circular orbit, $D\sim 3M$ and it follows that $MB_0\gtrsim 0.1$, which clashes with the assumption $MB_0\ll 1$ (for a more complete discussion, see~\cite{Brito:2014nja}). Despite this caveat, the main argument of the paper holds at least from a purely theoretical perspective.

Furthermore, physically meaningful solutions to the radial equation satisfy the inner boundary condition
\begin{align}
\left. R\right|_{y\rightarrow-\infty}\sim e^{-i(\omega-m\Omega_\mathcal{H})y}\ ,
\end{align}
i.e. they behave as waves falling into (emanating from) the black hole when $w>m\Omega_\mathcal{H}$ ($0<w<m\Omega_\mathcal{H}$).

\section{\label{sec:level3}Stationary scalar clouds}

When the scalar field's phase angular velocity is a natural multiple of the black hole's angular velocity, i.e.
\begin{align}
w=m\Omega_\mathcal{H}=-\frac{2mQB_0}{r_+}+\mathcal{O}({B_0^3})\ ,
\label{eq:3.1}
\end{align}
bound states, known as \textit{stationary clouds}, are found. Equation~\eqref{eq:3.1} is called \textit{synchronisation condition} and does depend on the scalar field's effective mass, $\mu_\text{eff}=\sqrt{m^2B_0^2}$. The ratio $|w/\mu_\text{eff}|=2|Q|/r_+$ is independent of $B_0$ and its absolute value is smaller than or equal to~{$2$}. Since it was assumed that $\sgn(w)=\sgn(B_0)=+1$, the synchronisation condition dictates that the bound states satisfy $\sgn(mQ)=-1$. 

The synchronisation occurs in one--dimensional subsets of the two--dimensional parameter space of Reissner--Nordst\"{o}m--Melvin black holes, described by $\{M,Q\}$. These subsets -- known as~\textit{existence lines} -- are disjoint and can be labeled with a set of three ``quantum'' numbers: the number of nodes in the radial direction~\footnote{The number of nodes in the radial direction does not include the node at $r=r_0$ when~$\zeta=0$ (Dirichlet boundary condition).} $n$, the orbital/total angular momentum $\ell$ and the azimuthal harmonic index $m$. These states will be labeled with $\ket{n,\ell,m}$.

In the following, stationary scalar clouds around RNM black holes are obtained both (semi--)analytically and numerically. The existence lines will be plotted in the $(M,Q)$--plane normalized to the magnetic field strength $B_0$. 

\subsection{\label{sec:level3.1}Analytical approach}

The eigenvalue problem at hand can be solved using the matched asymptotic expansion method (see, \textit{e.g.},~\cite{Cardoso:2008kj}), i.e. constructing approximations to the solutions of \eqref{eq:3.3a} that separately satisfy the inner and outer boundary conditions. The interval $r\in[r_+,r_0]$ is thus split into two: (i) the inner region, $r-r_+\ll\lambda_c$, where $\lambda_c=\mu_\text{eff}^{-1}\leq r_+/(m|Q|B_0)$ is the scalar field's Compton wavelength; inspection shows that $\lambda_c\gg M$; and (ii) the outer region, $r-r_+\gg M$.  The inner and outer expansions are then matched in the overlap region, where both conditions  can  hold simultaneously, defined by $M\ll r-r_+\ll\lambda_c$.

\subsubsection{Outer region}

The outer region is well--defined only if the outer boundary is sufficiently far from the black hole, i.e. as long as $r_0\gg M$. Given that $Q^2\leq M^2$, one can take~$\Delta\sim r^2$. Besides, if~$r^2\gg|2mQB_0/w|$, then $K\sim wr^2$. When the syncrhonization condition~\eqref{eq:3.1} holds, the latter approximation is equivalent to $r\gg r_+$, which is consistent with $r-r_+\gg M$.

The radial equation~\eqref{eq:3.3a} then reduces to that of a massless scalar field perturbation with phase angular velocity defined by {$\varpi^2\equiv w^2-\mu_\text{eff}^2=m^2B_0^2(4Q^2/r_+^2-1)$} and angular momentum $\ell$ in Minkowski spacetime~\footnote{Alternatively, one could say that Eq.~\eqref{eq:4.1.1} describes a scalar field with mass $\sqrt{m^2B^2}$, phase angular velocity $m\Omega_\mathcal{H}$ and angular momentum $\ell$.},
\begin{align}
\frac{\D^2}{\D r^2}(rR_+)+\left[\varpi^2-\frac{\ell(\ell+1)}{r^2}\right](rR_+)=0\ ,
\label{eq:4.1.1}
\end{align}
where~$R_+(r)\equiv\lim_{r\rightarrow r_0}R(r)$. The general solution is
\begin{align}
R_+(r)=\alpha_+ j_\ell(\varpi r)+\beta_+ y_\ell(\varpi r)\ ,
\label{eq:4.1.2}
\end{align}
where $j_\ell$ and $y_\ell$ are the spherical Bessel functions of the first and second kinds, respectively, and $\alpha_+,\beta_+\in\mathbb{C}$. {For sufficiently large~$r$, the spherical Bessel functions are a linear combination of ingoing and outgoing waves if $\varpi$ is real, i.e. if $w^2>\mu_\text{eff}^2$}. The Robin boundary condition~\eqref{eq:4.3} fixes the quotient
\begin{align}
\gamma\equiv\frac{\beta_+}{\alpha_+}=\left.\left[-\frac{j_\ell(\varpi r)+\tan(\zeta)j_\ell'(\varpi r)}{y_\ell(\varpi r)+\tan(\zeta)y_\ell'(\varpi r)}\right]\right|_{r=r_0}\ .
\label{eq:4.1.3}
\end{align}
The small--$r$ behavior of the asymptotic solution~\eqref{eq:4.1.2} is
\begin{align}
R_+(r)\sim\alpha_+\frac{(\varpi r)^\ell}{(2\ell+1)!!}-\beta_+\frac{(2\ell-1)!!}{(\varpi r)^{\ell+1}}\ .
\end{align}

\subsubsection{Inner region}

Near the outer horizon, the radial equation~\eqref{eq:3.3a} reduces to
\begin{align}
\frac{\D}{\D r}\left(\Delta\frac{\D R_-}{\D r}\right)-\ell(\ell+1)R_-=0\ ,
\label{eq:4.1.4}
\end{align}
where~$R_-(r)\equiv\lim_{r\rightarrow r_+}R(r)$.
Introducing the radial coordinate~$z\equiv(r-r_+)/(r-r_-)$ and defining~$R_-(z)=(1-z)^{\ell+1}F(z)$, one can bring the radial equation~\eqref{eq:4.1.4} into the form
\begin{align}
z(1-z)\frac{\D^2F}{\D z^2}+\left[c-(a+b+1)z\right]\frac{\D F}{\D z}-abF=0\ ,
\label{eq:4.1.5}
\end{align}
with $a=b\equiv\ell+1$ and $c\equiv1$. Equation~\eqref{eq:4.1.5} is a Gaussian hypergeometric equation, which has three regular singular points: $z=0,1,\infty$. The most general solution is~\cite{mabramowitz64:handbook,Herdeiro:2016tmi}
\begin{align}
F(z)=
&\alpha_-F(a,a;1;z)\nonumber\\
&+\beta_-\left[F(a,a;1;z)\log z+2\sum_{j=1}^{+\infty}f_{(j)}z^j\right]\ ,
\label{eq:4.1.6}
\end{align}
where
\begin{align*}
f_{(j)}=\left[\frac{(a)_j}{j!}\right]^2\left[\psi(a+j)-\psi(a)-\psi(j+1)+\psi(1)\right]
\end{align*}
and $(a)_j=\Gamma(a+j)/\Gamma(a)$ and $\psi$ is the digamma function. The second term in Eq.~\eqref{eq:4.1.6} diverges logarithmically as $z\rightarrow0$ ($r\rightarrow r_+$).  As the inner boundary condition must be regular, the constant $\beta_-$ must vanish. In terms of the radial function $R_-$, the solution thus reads
\begin{align*}
R_-(z)=\alpha_-\frac{(2\ell+1)!}{(\ell!)^2}\left[(-1)^{2\ell+1}R^{(\text{D})}(z)+\frac{R^{(\text{N})}(z)}{(\ell+1)^2}\right]\ ,
\end{align*}
where
\begin{align*}
R_-^{(\text{D})}(z)&=(1-z)^{\ell+1}F(\ell+1,\ell+1;2\ell+2;1-z)\ ,\\
R_-^{(\text{N})}(z)&=(1-z)^{-\ell}F(-\ell,-\ell;-2\ell;1-z)\ .
\end{align*}
When $r\gg M$, $z\sim 1$ and $(1-z)\sim(r_+-r_-)/r$, meaning that
\begin{align*}
R_-^{(\text{D})}(z)&\sim(r_+-r_-)^{\ell+1}r^{-\ell-1}\ ,\\
R_-^{(\text{N})}(z)&\sim(r_+-r_-)^{-\ell}r^{\ell}\ .
\end{align*}

\subsubsection{Matching}
\label{secmat}

It is clear that the larger--$r$ behavior of the asymptotic solution~$R_-$ exhibits the same dependence on $r$ as the small--$r$ behavior of the asymptotic solution $R_+$. Matching the two solutions, one gets
\begin{align}
\gamma=\frac{(\ell+1)^2}{(2\ell+1)!!(2\ell-1)!!}\left[\varpi(r_+-r_-)\right]^{2\ell+1}\ .
\label{eq:4.1.11}
\end{align}
Using Eq.~\eqref{eq:4.1.3}, one finally obtains
\begin{align}
&\tan(\zeta)=-\frac{j_\ell(\varpi r_0)+\gamma y_\ell(\varpi r_0)}{j_\ell'(\varpi r_0)+\gamma y_\ell'(\varpi r_0)}\ ,
\label{eq:4.1.12}
\end{align}
which establishes the existence condition for stationary scalar clouds around (non--extremal) RNM black holes. These exist as long as the field perturbation has a radial oscillatory character and therefore can satisfy a Robin boundary condition at $r_0B_0\sim 1$. This requirement is met provided that $\varpi$ is real, i.e. if
\begin{align}
w^2>\mu_\text{eff}^2~\Leftrightarrow~
\frac{4Q^2}{r_+^2}>1
~\implies~\frac{Q^2}{M^2}>\frac{16}{25}\ ,
\end{align}
or $|Q/M|\in(0.8,1.0)$, where $\sgn(Q)=\pm1$ for $\sgn(m)=\mp 1$ so that $\sgn(w)=+1$. Note that this restriction on the specific electric charge is a by--product of the proportionality between $w=m\Omega_\mathcal{H}$ and $\mu_\text{eff}$.

\subsection{\label{sec:level3.2}Numerical approach}

Stationary clouds can also be found by solving numerically the coupled equations~\eqref{eq:3.3a}{--\eqref{eq:3.3b}}. For that purpose, it is convenient to replace the mass~$M$ by the outer horizon radius~$r_+$ and work with the dimensionless quantities $\{r_+B_0,QB_0,\Omega_\mathcal{H}/B_0\}$. To impose the correct inner boundary condition the radial function may be written as a series expansion around $r=r_+$~\cite{Pani:2013pma},
\begin{align}
\left.R\right|_{r\rightarrow r_+}\sim\sum_{j=0}^{+\infty}a_{(j)}(r-r_+)^j\ .
\label{eq:4.3}
\end{align} 
The coefficients $\{a_{(j)}\}_{j>0}$ are obtained by plugging~\eqref{eq:4.3} into~\eqref{eq:3.3a}, writing the resulting equation in powers of $(r-r_+)$ and setting the coefficient of each power separately equal to zero. The resulting system of equations must then be solved for $\{a_{(j)}\}_{j>0}$ in terms of $a_{(0)}$. The latter is set to~$1$ without loss of generality. The coefficients $\{a_{(j)}\}_{j>0}$ depend on the black hole's parameters $\{r_+,Q\}$, the Killing eigenvalue $m$ and the separation constant $\lambda$. {Instead of solving the angular equation~\eqref{eq:3.3b}, one approximates the latter by~$\ell(\ell+1)$, which is accurate enough if $mQB_0\ll 1$. Since $Q^2\leq M^2$ and $MB_0\ll MB=1$, the approximation is valid for moderate values of $m$.}

The parameters $\{r_+,\ell,m\}$ are assigned fixed values.  By virtue of the regular singular point at $r=r_+$, Eq.~\eqref{eq:3.3a} must be integrated from $r=r_+(1+\delta)$, with $\delta\ll 1$, to $r=r_0$, where $r_0$ is the outer boundary radial coordinate. A simple shooting method finds  the $Q$--values for which the numerical solutions satisfy a Robin boundary condition at $r=r_0$. 

\subsection{\label{sec:level3.3}Existence lines}

\begin{figure}[t!]
\includegraphics[width=\columnwidth]{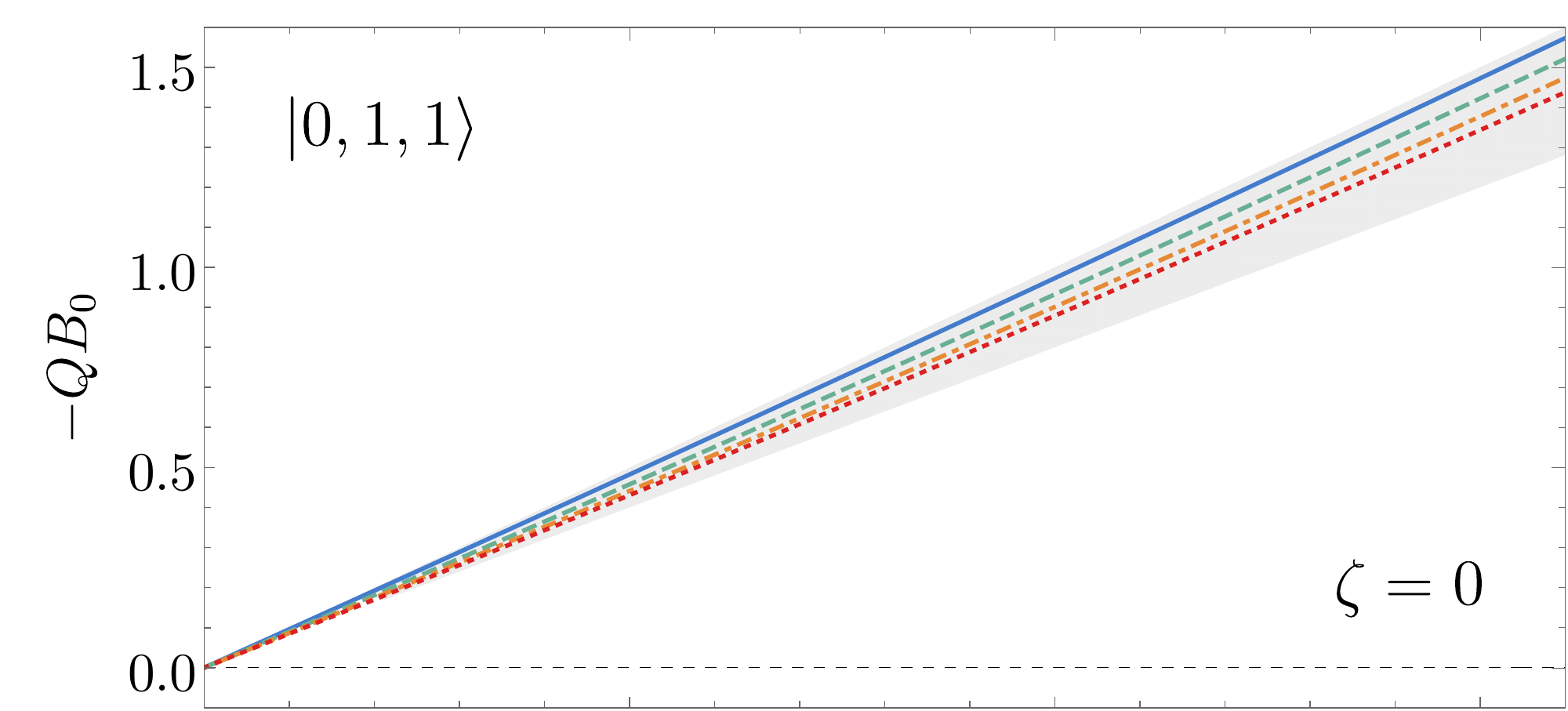}
\includegraphics[width=\columnwidth]{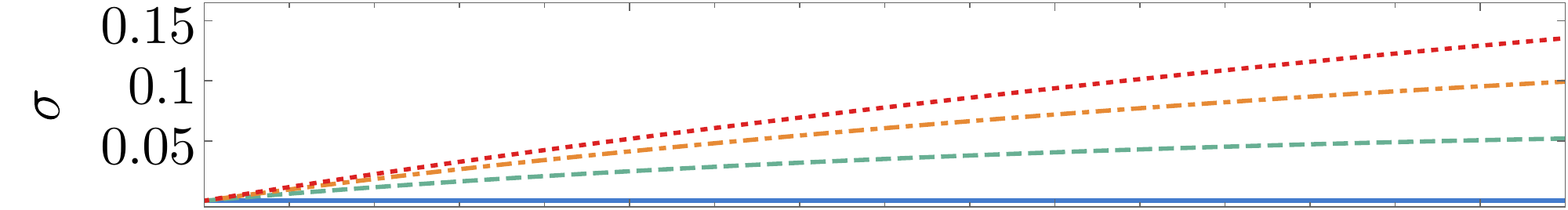}
\includegraphics[width=\columnwidth]{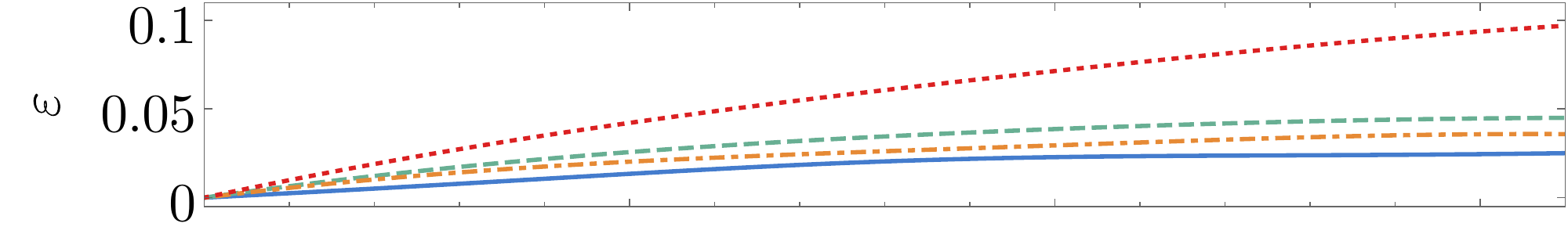}
\includegraphics[width=\columnwidth]{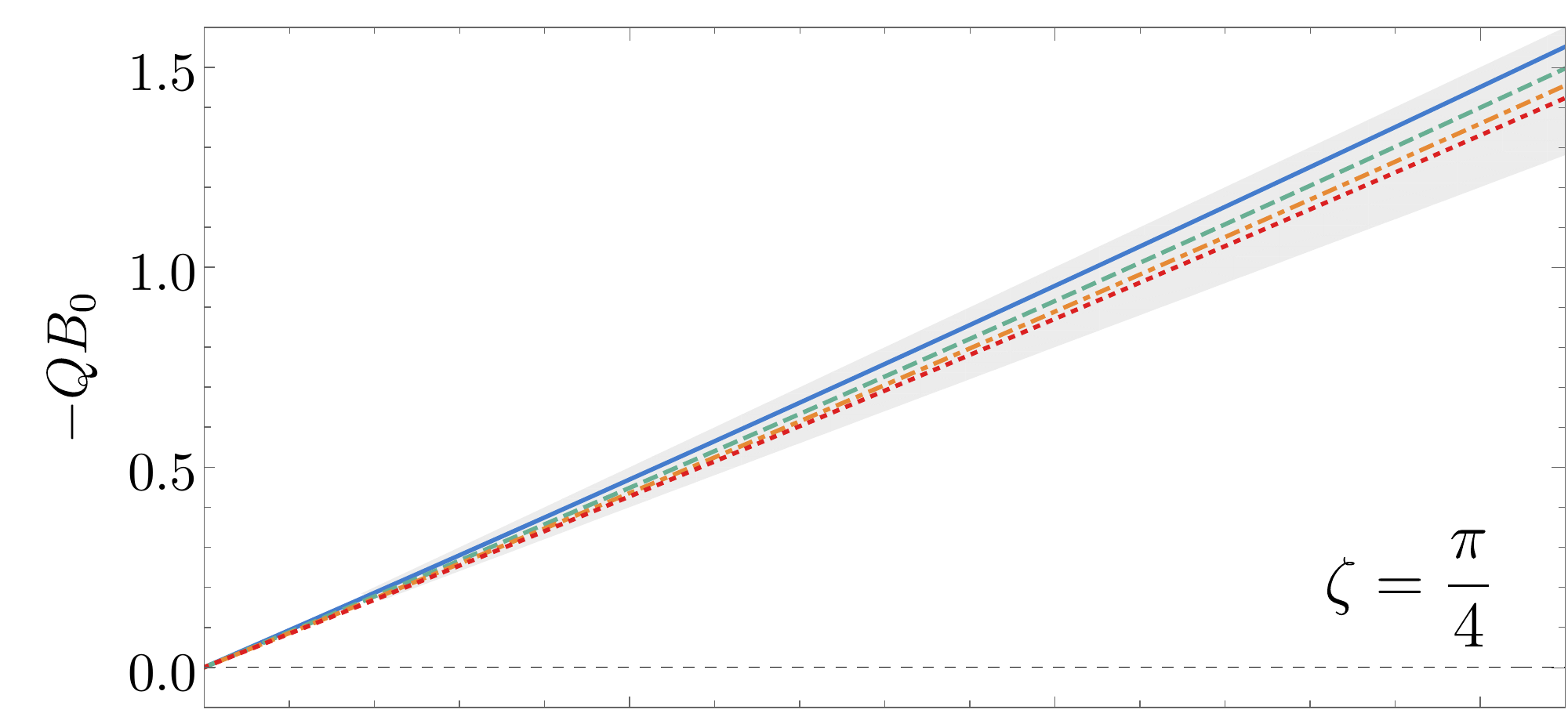}
\includegraphics[width=\columnwidth]{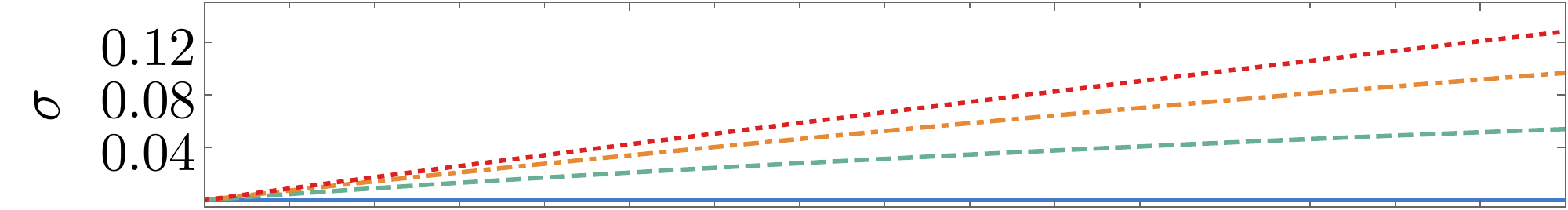}
\includegraphics[width=\columnwidth]{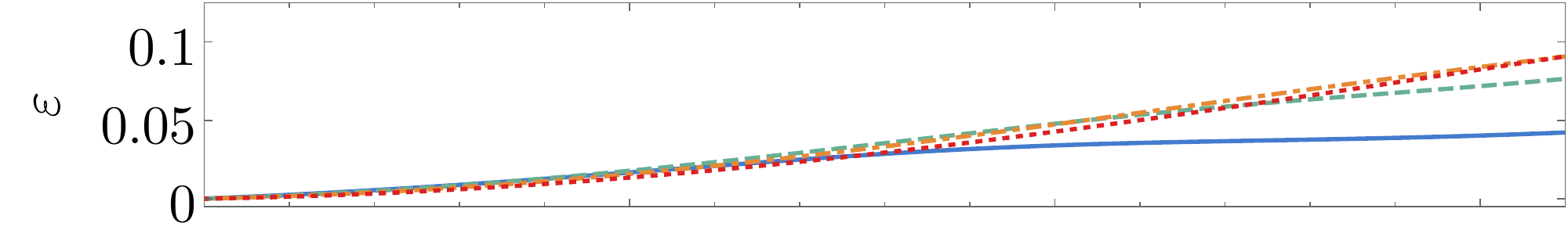}
\includegraphics[width=\columnwidth]{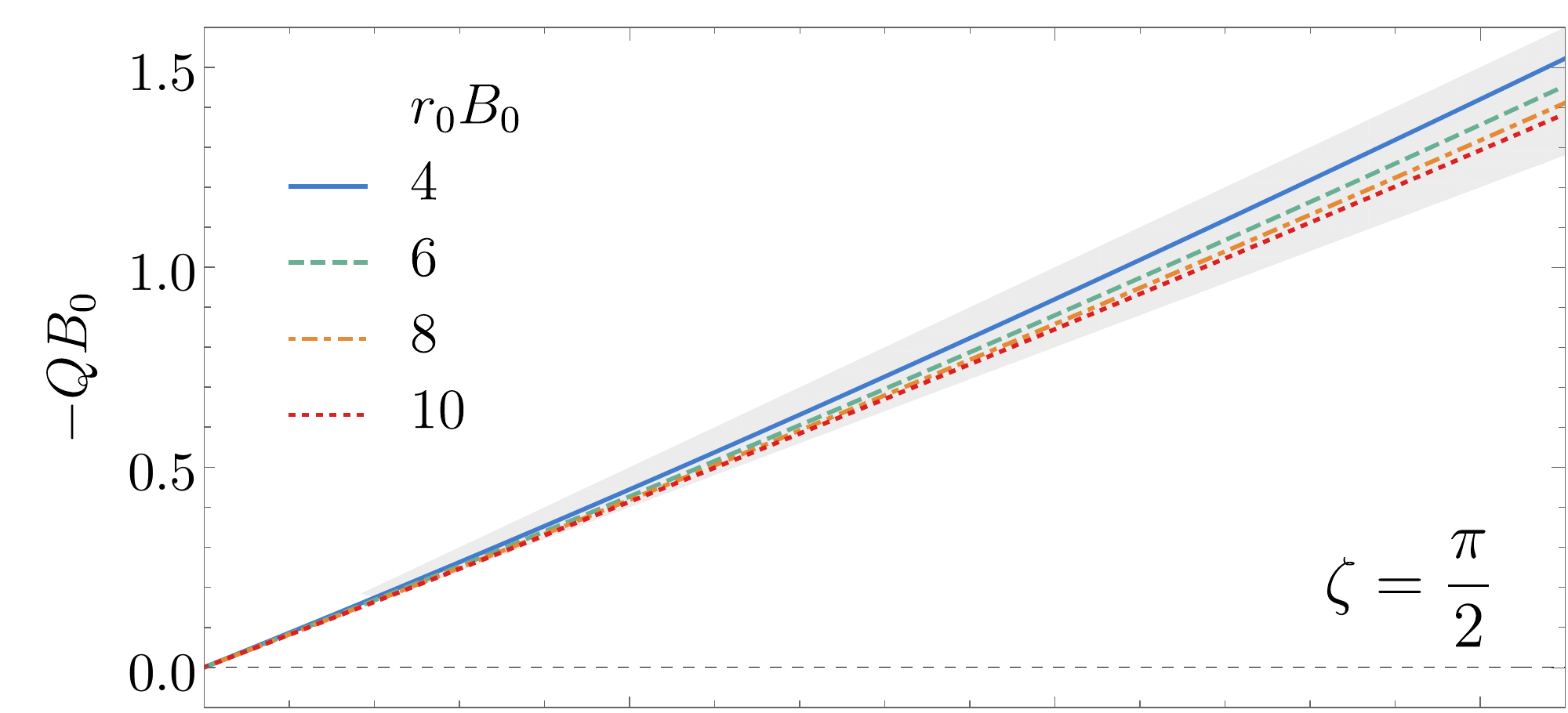}
\includegraphics[width=\columnwidth]{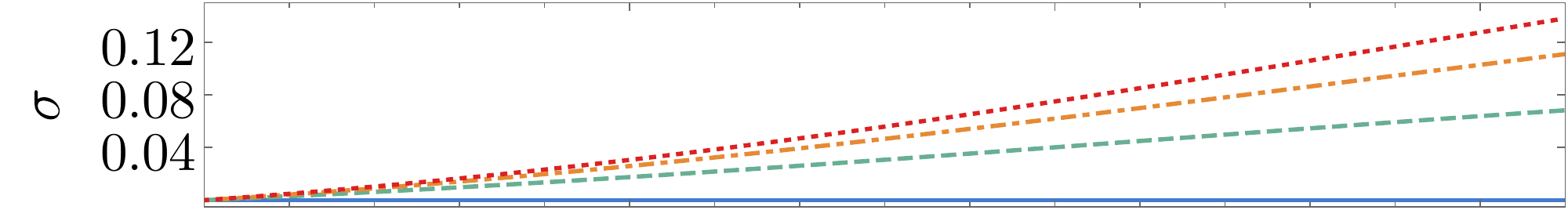}
\includegraphics[width=\columnwidth]{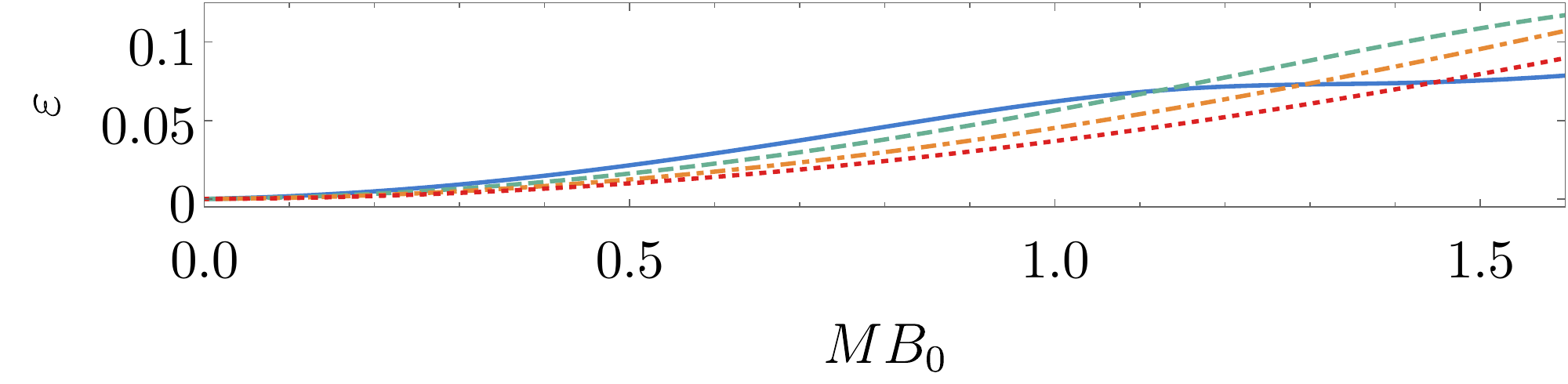}
\caption{Stationary scalar clouds $\ket{n,\ell,m}=\ket{0,1,1}$ around Reissner--Nordstr\"{o}m black holes embedded in a uniform axial magnetic field of strength $B_0$, for different Robin boundary conditions, parametrized by $\zeta$, at the outer boundary $r_0$. Fixing $MB_0$, }
\label{fig:1}
\end{figure}

\autoref{fig:1} displays the (numerical) existence lines for stationary clouds $\ket{0,1,1}$ with $r_0B_0\in\{4,6,8,10\}$ and~$\zeta\in\{0,\frac{\pi}{2},\frac{\pi}{4}\}$. The shaded bands represent the allowed regions of the parameter space for the existence of bound states. The upper boundary, defined by $Q^2=M^2$, corresponds to the \textit{extremal line}. The RNM black holes in the lower boundary satisfy $Q^2=0.64M^2$, in accordance with the conclusion at the end of section~\ref{secmat}.

The panels below the main plots show the absolute difference $\sigma$ between each existence line and that corresponding to $r_0B_0=4$ and the absolute difference~$\varepsilon$ between the \textit{numerical} and \textit{analytical} existence lines. As expected, given that the analytical condition~\eqref{eq:4.1.12} is valid when~$MB_0\ll1$,~$\varepsilon\rightarrow0$ as~$MB_0\rightarrow0$.

All existence lines lie within the shaded bands. Also, they converge to $(M,Q)=(0,0)$, i.e. $\sigma\rightarrow0$ as~$MB_0\rightarrow0$, which is in agreement with the expectation that scalar field perturbations cannot attain stationary equilibrium with respect to {asymptotically--Melvin black holes}. Fixing $MB_0$, as the region of influence of the magnetic field decreases, i.e. as $r_0B_0$ decreases, the Coulomb energy of the black hole supporting the stationary cloud increases. Vaster clouds thus require lower angular velocities so that they do not collapse into the black hole. Also, there is an overall decrease in the Coloumb energy as $\zeta$ varies continuously from $0$ (Dirichlet boundary condition) to $\frac{\pi}{2}$ (Neumann boundary condition).

\begin{figure}[h]
\includegraphics[width=\columnwidth]{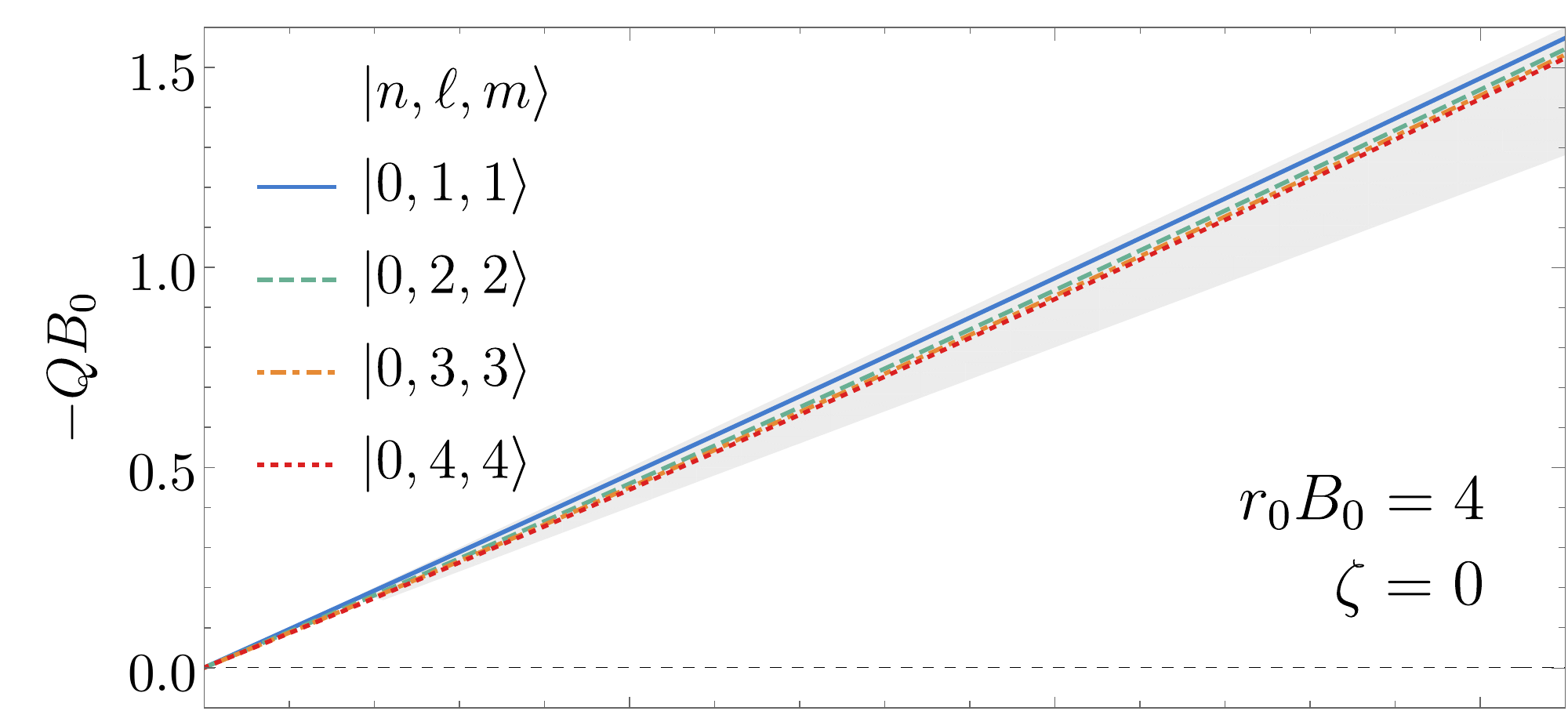}
\includegraphics[width=\columnwidth]{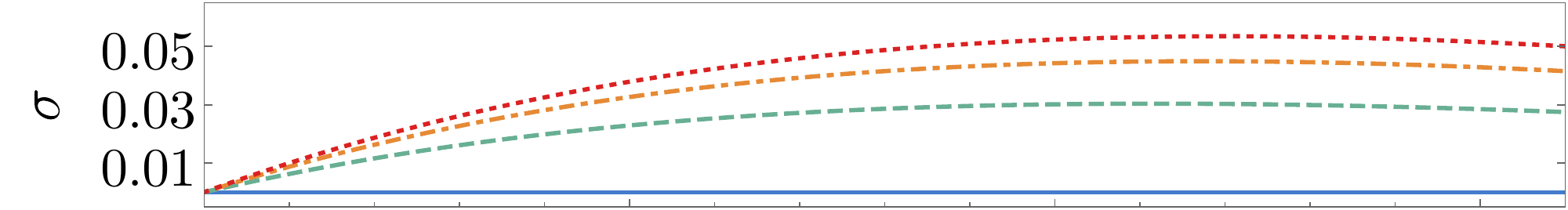}
\includegraphics[width=\columnwidth]{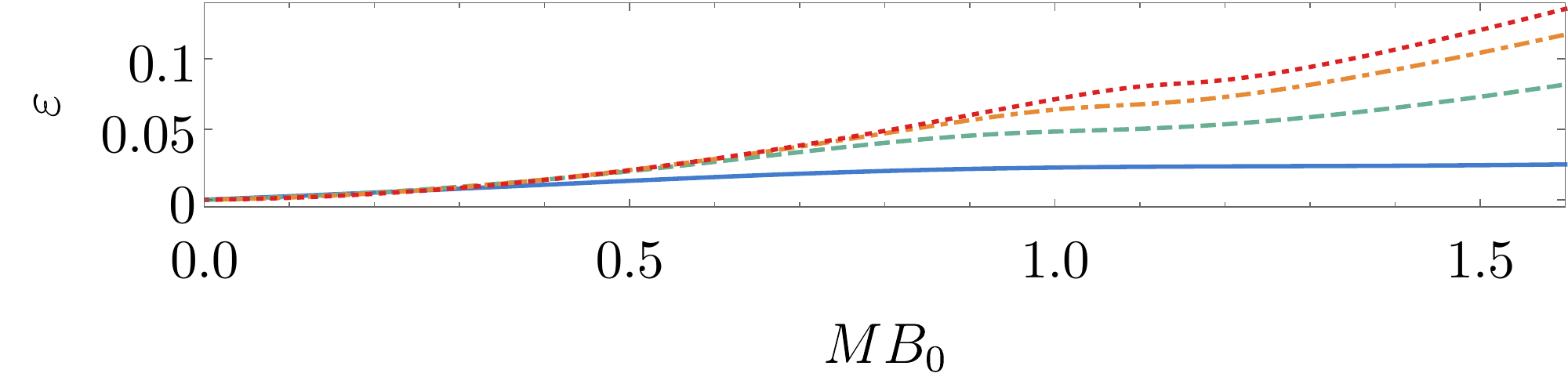}
\caption{Stationary scalar clouds $\ket{n,\ell,m}=\ket{0,\ell,\ell}$ around Reissner--Nordstr\"{o}m black holes embedded in a uniform axial magnetic field of strength $B_0$ and satisfying a Dirichlet boundary condition ($\zeta=0$) at $r_0B_0=4$.}
\label{fig:2}
\end{figure}

The existence lines for the states~$\ket{0,\ell,m}$ with $\ell=m=1,\ldots,4$, $r_0B_0=4$ and $\zeta=0$ are plotted in~\autoref{fig:2}.  These approach the extremal line as $\ell=m$ decreases, a trend already noticed in previous works (see, \textit{e.g.},~\cite{Benone:2014ssa}).  

\begin{figure}[h]
\includegraphics[width=\columnwidth]{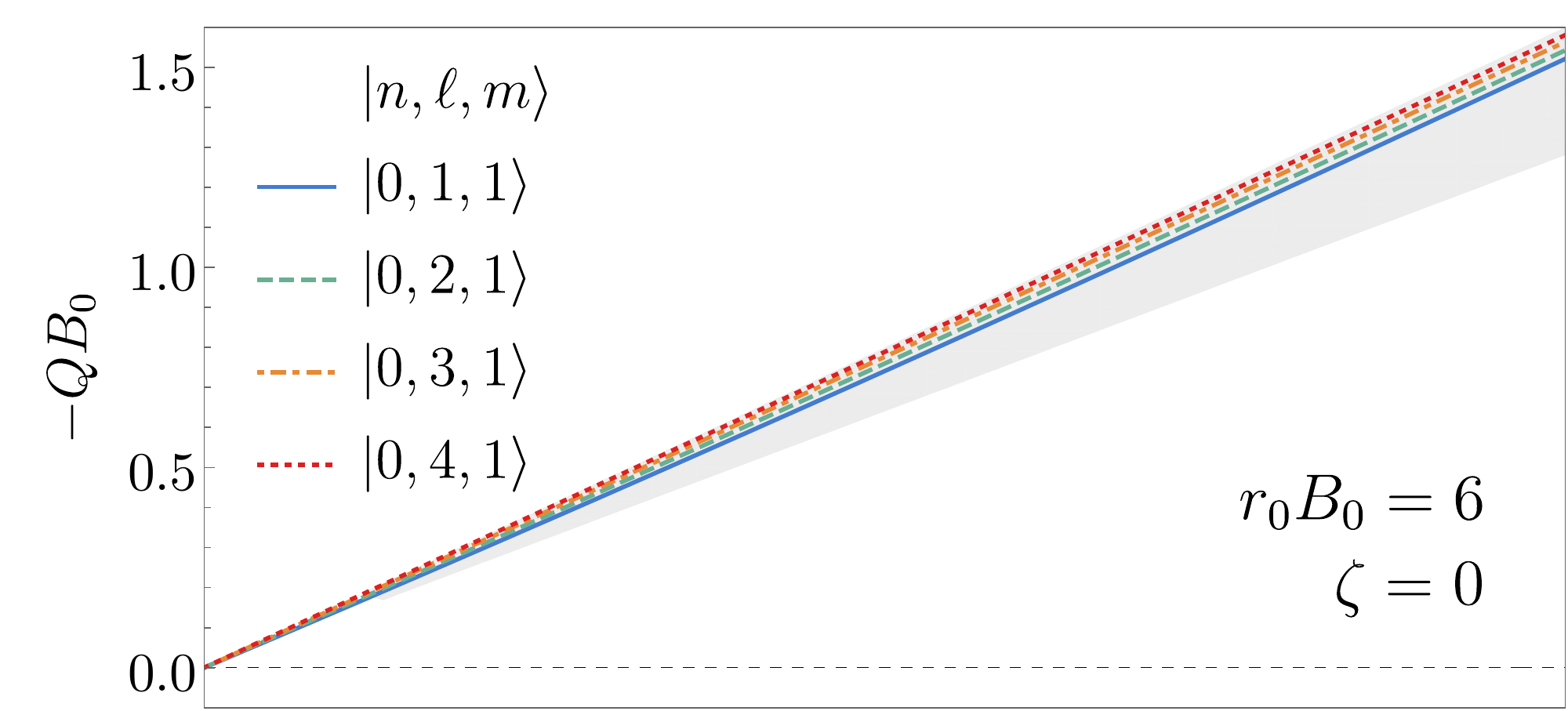}
\includegraphics[width=\columnwidth]{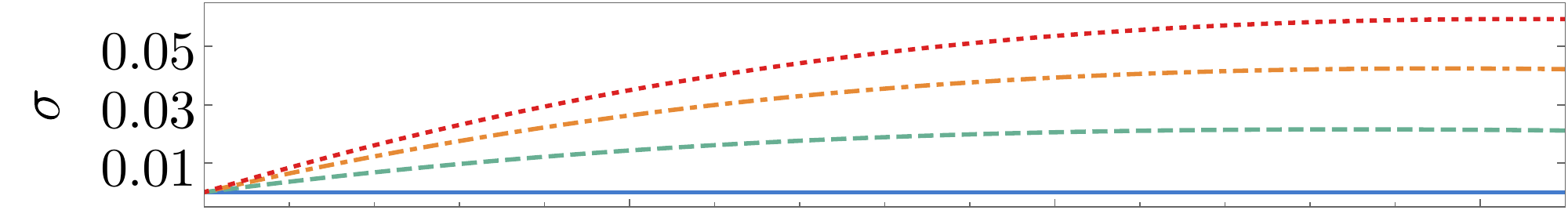}
\includegraphics[width=\columnwidth]{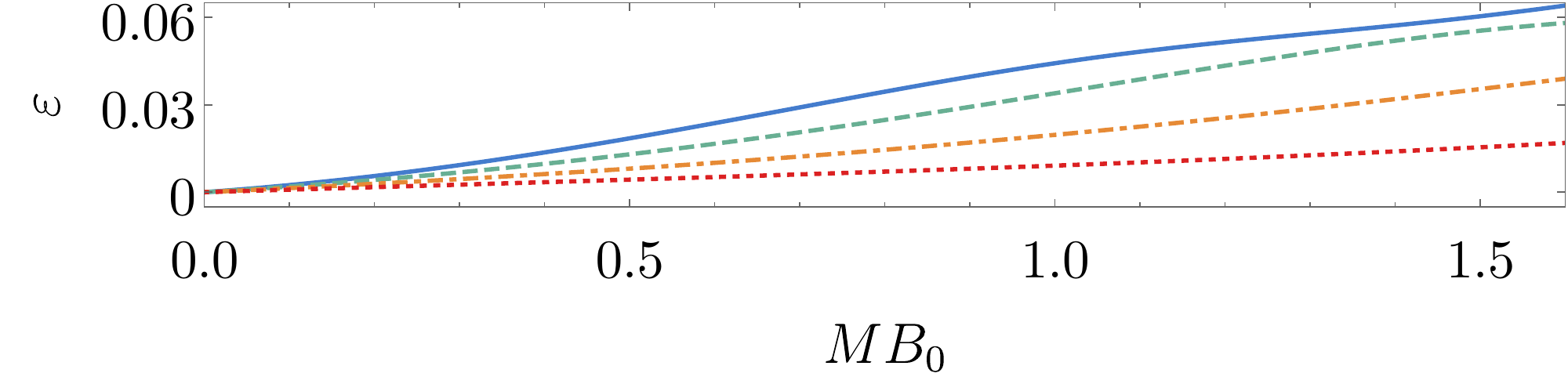}
\caption{Stationary scalar clouds $\ket{n,\ell,m}=\ket{0,\ell,1}$, with $\ell=1,\ldots,4$, around Reissner--Nordstr\"{o}m black holes embedded in a uniform axial magnetic field of strength $B_0$ and satisfying a Dirichlet boundary condition ($\zeta=0$) at $r_0B_0=6$.}
\label{fig:3}
\end{figure}

The impact of the orbital angular momentum~$\ell$ is enlightned in~\autoref{fig:3}, in which the existence lines for the states~$\ket{0,\ell,1}$ with $\ell=1,\ldots,4$, $r_0B_0=6$ and $\zeta=0$ are shown. As~$\ell$ increases, so does $|Q/M|$, which suggests that stationary clouds~$\ket{0,\ell,1}$ with $\ell>1$ are more energetic than~$\ket{0,1,1}$. 

\section{\label{sec:level4}Conclusion}

The RNM black hole stands out as a toy model for a rotating black hole immersed in an external axial magnetic field. In fact, it is the simplest stationary  (but not static) solution of Einstein--Maxwell equations asymptotically resembling the magnetic Melvin universe. Frequently overlooked due to its astrophysical irrelevance, it is still worth studying as it may offer some insights into the interaction of black holes with magnetic fields.

The present paper aimed precisely to explore the interplay between bosonic fields and black holes when permeated by a uniform magnetic field. It was shown in particular that RNM black holes support synchronized scalar field configurations known as stationary clouds. They are somehow akin to atomic orbitals of the hydrogen atom in quantum mechanics in that they are both described by quantum number. In effect, stationary clouds are characterized by the number of nodes in the radial direction, $n$, the orbital angular momentum, $\ell$, and the azimuthal harmonic index, $m$, which labels the projection of 
the orbital angular momentum along the direction of the magnetic field. 

It is now well known that stationary equilibrium is possible whenever a bosonic field at the threshold of superradiant instabilities (i.e. obeying the so--called syncrhonization condition) is confined in the black hole's vicinity. The confinement mechanism (either natural or artificial) creates a potential barrier which may prevent the field from escaping to infinity. As a result, infinitely long--lived configurations arise. For example, a massive bosonic field can form such stationary clouds around Kerr black holes -- with the field's mass providing a natural confinement mechanism. So does a massless charged scalar field in a cavity enclosing a Reissner--Nordstr\"{o}m black hole -- with the boundary of the cavity, a reflective mirror, sourcing an artificial confinement mechanism~\cite{Herdeiro:2013pia}. The properties of both equilibrium configurations are similar despite minor qualitative differences. 

Additionally worth mentioning is the fact that, in two previous examples, the occurrence of superradiance does not rely on the existence of a confining environment; one could say that the two ingredients are added separately. However, in the setup under consideration, the magnetic field of the RNM black hole is responsible not only for developing an ergoregion and hence trigger superradiant phenomena but also for making low--frequency fields acquire an effective mass and thus be trapped, allowing the formation of stationary clouds. In view of this, it does not come as a surprise that both the black hole's angular velocity~$\Omega_\mathcal{H}$ and the field's effective mass~$\mu_\text{eff}$ -- synonyms for superradiance and confinement, respectively -- depend on $B_0$.

Lastly, a by--product of considering the RNM black hole was the realization that the quotient $m\Omega_\mathcal{H}/\mu_\text{eff}$ is a function of the black hole's specific electric charge $Q/M$ only. Consequently, the condition for the existence of bound states constrains the values of $Q/M$ for which stationary clouds can exist. 

\section{Acknowledgements}

This work has been supported  by the Center for Astrophysics and Gravitation (CENTRA) and by the Center for Research and Development in Mathematics and Applications (CIDMA) through the Portuguese Foundation for Science and Technology (FCT -- Funda\c{c}\~ao para a Ci\^encia e a Tecnologia), references UIDB/00099/2020, UIDB/04106/2020 and UIDP/04106/2020. The authors acknowledge support  from the projects PTDC/FIS-OUT/28407/2017, CERN/FIS-PAR/0027/2019 and PTDC/FIS-AST/3041/2020. N. M. Santos is supported by the FCT grant
SFRH/BD/143407/2019. This work has further been supported by  the  European  Union's  Horizon  2020  research  and  innovation  (RISE) program H2020-MSCA-RISE-2017 Grant No.~FunFiCO-777740. The authors would like to acknowledge networking support by the COST Action CA16104.

\bibliography{apssamp}

\providecommand{\noopsort}[1]{}\providecommand{\singleletter}[1]{#1}%
\begin{thebibliography}{50}%
\makeatletter
\providecommand \@ifxundefined [1]{%
 \@ifx{#1\undefined}
}%
\providecommand \@ifnum [1]{%
 \ifnum #1\expandafter \@firstoftwo
 \else \expandafter \@secondoftwo
 \fi
}%
\providecommand \@ifx [1]{%
 \ifx #1\expandafter \@firstoftwo
 \else \expandafter \@secondoftwo
 \fi
}%
\providecommand \natexlab [1]{#1}%
\providecommand \enquote  [1]{``#1''}%
\providecommand \bibnamefont  [1]{#1}%
\providecommand \bibfnamefont [1]{#1}%
\providecommand \citenamefont [1]{#1}%
\providecommand \href@noop [0]{\@secondoftwo}%
\providecommand \href [0]{\begingroup \@sanitize@url \@href}%
\providecommand \@href[1]{\@@startlink{#1}\@@href}%
\providecommand \@@href[1]{\endgroup#1\@@endlink}%
\providecommand \@sanitize@url [0]{\catcode `\\12\catcode `\$12\catcode
  `\&12\catcode `\#12\catcode `\^12\catcode `\_12\catcode `\%12\relax}%
\providecommand \@@startlink[1]{}%
\providecommand \@@endlink[0]{}%
\providecommand \url  [0]{\begingroup\@sanitize@url \@url }%
\providecommand \@url [1]{\endgroup\@href {#1}{\urlprefix }}%
\providecommand \urlprefix  [0]{URL }%
\providecommand \Eprint [0]{\href }%
\providecommand \doibase [0]{http://dx.doi.org/}%
\providecommand \selectlanguage [0]{\@gobble}%
\providecommand \bibinfo  [0]{\@secondoftwo}%
\providecommand \bibfield  [0]{\@secondoftwo}%
\providecommand \translation [1]{[#1]}%
\providecommand \BibitemOpen [0]{}%
\providecommand \bibitemStop [0]{}%
\providecommand \bibitemNoStop [0]{.\EOS\space}%
\providecommand \EOS [0]{\spacefactor3000\relax}%
\providecommand \BibitemShut  [1]{\csname bibitem#1\endcsname}%
\let\auto@bib@innerbib\@empty
\bibitem [{\citenamefont {Abbott}\ \emph {et~al.}(2019)\citenamefont {Abbott}
  \emph {et~al.}}]{LIGOScientific:2018mvr}%
  \BibitemOpen
  \bibfield  {author} {\bibinfo {author} {\bibfnamefont {B.}~\bibnamefont
  {Abbott}} \emph {et~al.} (\bibinfo {collaboration} {LIGO Scientific,
  Virgo}),\ }\href {\doibase 10.1103/PhysRevX.9.031040} {\bibfield  {journal}
  {\bibinfo  {journal} {Phys. Rev. X}\ }\textbf {\bibinfo {volume} {9}},\
  \bibinfo {pages} {031040} (\bibinfo {year} {2019})},\ \Eprint
  {http://arxiv.org/abs/1811.12907} {arXiv:1811.12907 [astro-ph.HE]}
  \BibitemShut {NoStop}%
\bibitem [{\citenamefont {Cardoso}\ \emph {et~al.}(2018)\citenamefont
  {Cardoso}, \citenamefont {Ikeda}, \citenamefont {Moore},\ and\ \citenamefont
  {Yoo}}]{Cardoso:2018nkg}%
  \BibitemOpen
  \bibfield  {author} {\bibinfo {author} {\bibfnamefont {V.}~\bibnamefont
  {Cardoso}}, \bibinfo {author} {\bibfnamefont {T.}~\bibnamefont {Ikeda}},
  \bibinfo {author} {\bibfnamefont {C.~J.}\ \bibnamefont {Moore}}, \ and\
  \bibinfo {author} {\bibfnamefont {C.-M.}\ \bibnamefont {Yoo}},\ }\href
  {\doibase 10.1103/PhysRevD.97.084013} {\bibfield  {journal} {\bibinfo
  {journal} {Phys. Rev. D}\ }\textbf {\bibinfo {volume} {97}},\ \bibinfo
  {pages} {084013} (\bibinfo {year} {2018})},\ \Eprint
  {http://arxiv.org/abs/1803.03271} {arXiv:1803.03271 [gr-qc]} \BibitemShut
  {NoStop}%
\bibitem [{\citenamefont {Frederiks}\ \emph {et~al.}(2013)\citenamefont
  {Frederiks} \emph {et~al.}}]{Frederiks:2013cga}%
  \BibitemOpen
  \bibfield  {author} {\bibinfo {author} {\bibfnamefont {D.}~\bibnamefont
  {Frederiks}} \emph {et~al.},\ }\href {\doibase 10.1088/0004-637X/779/2/151}
  {\bibfield  {journal} {\bibinfo  {journal} {Astrophys. J.}\ }\textbf
  {\bibinfo {volume} {779}},\ \bibinfo {pages} {151} (\bibinfo {year}
  {2013})},\ \Eprint {http://arxiv.org/abs/1311.5734} {arXiv:1311.5734
  [astro-ph.HE]} \BibitemShut {NoStop}%
\bibitem [{\citenamefont {Abbott}\ \emph {et~al.}(2020)\citenamefont {Abbott}
  \emph {et~al.}}]{Abbott:2020niy}%
  \BibitemOpen
  \bibfield  {author} {\bibinfo {author} {\bibfnamefont {R.}~\bibnamefont
  {Abbott}} \emph {et~al.} (\bibinfo {collaboration} {LIGO Scientific,
  Virgo}),\ }\href@noop {} {\  (\bibinfo {year} {2020})},\ \Eprint
  {http://arxiv.org/abs/2010.14527} {arXiv:2010.14527 [gr-qc]} \BibitemShut
  {NoStop}%
\bibitem [{\citenamefont {Raaijmakers}\ \emph {et~al.}(2020)\citenamefont
  {Raaijmakers} \emph {et~al.}}]{Raaijmakers:2019dks}%
  \BibitemOpen
  \bibfield  {author} {\bibinfo {author} {\bibfnamefont {G.}~\bibnamefont
  {Raaijmakers}} \emph {et~al.},\ }\href {\doibase 10.3847/2041-8213/ab822f}
  {\bibfield  {journal} {\bibinfo  {journal} {Astrophys. J. Lett.}\ }\textbf
  {\bibinfo {volume} {893}},\ \bibinfo {pages} {L21} (\bibinfo {year}
  {2020})},\ \Eprint {http://arxiv.org/abs/1912.11031} {arXiv:1912.11031
  [astro-ph.HE]} \BibitemShut {NoStop}%
\bibitem [{\citenamefont {Vitale}\ and\ \citenamefont
  {Chen}(2018)}]{Vitale:2018wlg}%
  \BibitemOpen
  \bibfield  {author} {\bibinfo {author} {\bibfnamefont {S.}~\bibnamefont
  {Vitale}}\ and\ \bibinfo {author} {\bibfnamefont {H.-Y.}\ \bibnamefont
  {Chen}},\ }\href {\doibase 10.1103/PhysRevLett.121.021303} {\bibfield
  {journal} {\bibinfo  {journal} {Phys. Rev. Lett.}\ }\textbf {\bibinfo
  {volume} {121}},\ \bibinfo {pages} {021303} (\bibinfo {year} {2018})},\
  \Eprint {http://arxiv.org/abs/1804.07337} {arXiv:1804.07337 [astro-ph.CO]}
  \BibitemShut {NoStop}%
\bibitem [{\citenamefont {Olausen}\ and\ \citenamefont
  {Kaspi}(2014)}]{Olausen:2013bpa}%
  \BibitemOpen
  \bibfield  {author} {\bibinfo {author} {\bibfnamefont {S.}~\bibnamefont
  {Olausen}}\ and\ \bibinfo {author} {\bibfnamefont {V.}~\bibnamefont
  {Kaspi}},\ }\href {\doibase 10.1088/0067-0049/212/1/6} {\bibfield  {journal}
  {\bibinfo  {journal} {Astrophys. J. Suppl.}\ }\textbf {\bibinfo {volume}
  {212}},\ \bibinfo {pages} {6} (\bibinfo {year} {2014})},\ \Eprint
  {http://arxiv.org/abs/1309.4167} {arXiv:1309.4167 [astro-ph.HE]} \BibitemShut
  {NoStop}%
\bibitem [{\citenamefont {Frolov}\ and\ \citenamefont
  {Shoom}(2010)}]{Frolov:2010mi}%
  \BibitemOpen
  \bibfield  {author} {\bibinfo {author} {\bibfnamefont {V.~P.}\ \bibnamefont
  {Frolov}}\ and\ \bibinfo {author} {\bibfnamefont {A.~A.}\ \bibnamefont
  {Shoom}},\ }\href {\doibase 10.1103/PhysRevD.82.084034} {\bibfield  {journal}
  {\bibinfo  {journal} {Phys. Rev. D}\ }\textbf {\bibinfo {volume} {82}},\
  \bibinfo {pages} {084034} (\bibinfo {year} {2010})},\ \Eprint
  {http://arxiv.org/abs/1008.2985} {arXiv:1008.2985 [gr-qc]} \BibitemShut
  {NoStop}%
\bibitem [{\citenamefont {Brito}\ \emph {et~al.}(2015)\citenamefont {Brito},
  \citenamefont {Cardoso},\ and\ \citenamefont {Pani}}]{Brito:2015oca}%
  \BibitemOpen
  \bibfield  {author} {\bibinfo {author} {\bibfnamefont {R.}~\bibnamefont
  {Brito}}, \bibinfo {author} {\bibfnamefont {V.}~\bibnamefont {Cardoso}}, \
  and\ \bibinfo {author} {\bibfnamefont {P.}~\bibnamefont {Pani}},\ }\href
  {\doibase 10.1007/978-3-319-19000-6} {\emph {\bibinfo {title}
  {{Superradiance}: {Energy Extraction, Black-Hole Bombs and Implications for
  Astrophysics and Particle Physics}}}},\ Vol.\ \bibinfo {volume} {906}\
  (\bibinfo  {publisher} {Springer},\ \bibinfo {year} {2015})\ \Eprint
  {http://arxiv.org/abs/1501.06570} {arXiv:1501.06570 [gr-qc]} \BibitemShut
  {NoStop}%
\bibitem [{\citenamefont {Herdeiro}\ and\ \citenamefont
  {Radu}(2014)}]{Herdeiro:2014goa}%
  \BibitemOpen
  \bibfield  {author} {\bibinfo {author} {\bibfnamefont {C.~A.~R.}\
  \bibnamefont {Herdeiro}}\ and\ \bibinfo {author} {\bibfnamefont
  {E.}~\bibnamefont {Radu}},\ }\href {\doibase 10.1103/PhysRevLett.112.221101}
  {\bibfield  {journal} {\bibinfo  {journal} {Phys. Rev. Lett.}\ }\textbf
  {\bibinfo {volume} {112}},\ \bibinfo {pages} {221101} (\bibinfo {year}
  {2014})},\ \Eprint {http://arxiv.org/abs/1403.2757} {arXiv:1403.2757 [gr-qc]}
  \BibitemShut {NoStop}%
\bibitem [{\citenamefont {Herdeiro}\ \emph {et~al.}(2016)\citenamefont
  {Herdeiro}, \citenamefont {Radu},\ and\ \citenamefont
  {R\'narsson}}]{Herdeiro:2016tmi}%
  \BibitemOpen
  \bibfield  {author} {\bibinfo {author} {\bibfnamefont {C.}~\bibnamefont
  {Herdeiro}}, \bibinfo {author} {\bibfnamefont {E.}~\bibnamefont {Radu}}, \
  and\ \bibinfo {author} {\bibfnamefont {H.}~\bibnamefont {R\'narsson}},\
  }\href {\doibase 10.1088/0264-9381/33/15/154001} {\bibfield  {journal}
  {\bibinfo  {journal} {Class. Quant. Grav.}\ }\textbf {\bibinfo {volume}
  {33}},\ \bibinfo {pages} {154001} (\bibinfo {year} {2016})},\ \Eprint
  {http://arxiv.org/abs/1603.02687} {arXiv:1603.02687 [gr-qc]} \BibitemShut
  {NoStop}%
\bibitem [{\citenamefont {East}\ \emph {et~al.}(2014)\citenamefont {East},
  \citenamefont {Ramazano\u{g}lu},\ and\ \citenamefont
  {Pretorius}}]{East:2013mfa}%
  \BibitemOpen
  \bibfield  {author} {\bibinfo {author} {\bibfnamefont {W.~E.}\ \bibnamefont
  {East}}, \bibinfo {author} {\bibfnamefont {F.~M.}\ \bibnamefont
  {Ramazano\u{g}lu}}, \ and\ \bibinfo {author} {\bibfnamefont {F.}~\bibnamefont
  {Pretorius}},\ }\href {\doibase 10.1103/PhysRevD.89.061503} {\bibfield
  {journal} {\bibinfo  {journal} {Phys. Rev. D}\ }\textbf {\bibinfo {volume}
  {89}},\ \bibinfo {pages} {061503} (\bibinfo {year} {2014})},\ \Eprint
  {http://arxiv.org/abs/1312.4529} {arXiv:1312.4529 [gr-qc]} \BibitemShut
  {NoStop}%
\bibitem [{\citenamefont {Herdeiro}\ and\ \citenamefont
  {Radu}(2017)}]{Herdeiro:2017phl}%
  \BibitemOpen
  \bibfield  {author} {\bibinfo {author} {\bibfnamefont {C.~A.~R.}\
  \bibnamefont {Herdeiro}}\ and\ \bibinfo {author} {\bibfnamefont
  {E.}~\bibnamefont {Radu}},\ }\href {\doibase 10.1103/PhysRevLett.119.261101}
  {\bibfield  {journal} {\bibinfo  {journal} {Phys.\ Rev.\ Lett.}\ }\textbf
  {\bibinfo {volume} {119}},\ \bibinfo {pages} {261101} (\bibinfo {year}
  {2017})},\ \Eprint {http://arxiv.org/abs/1706.06597} {arXiv:1706.06597
  [gr-qc]} \BibitemShut {NoStop}%
\bibitem [{\citenamefont {Santos}\ \emph {et~al.}(2020)\citenamefont {Santos},
  \citenamefont {Benone}, \citenamefont {Crispino}, \citenamefont {Herdeiro},\
  and\ \citenamefont {Radu}}]{Santos:2020pmh}%
  \BibitemOpen
  \bibfield  {author} {\bibinfo {author} {\bibfnamefont {N.~M.}\ \bibnamefont
  {Santos}}, \bibinfo {author} {\bibfnamefont {C.~L.}\ \bibnamefont {Benone}},
  \bibinfo {author} {\bibfnamefont {L.~C.~B.}\ \bibnamefont {Crispino}},
  \bibinfo {author} {\bibfnamefont {C.~A.~R.}\ \bibnamefont {Herdeiro}}, \ and\
  \bibinfo {author} {\bibfnamefont {E.}~\bibnamefont {Radu}},\ }\href {\doibase
  10.1007/JHEP07(2020)010} {\bibfield  {journal} {\bibinfo  {journal} {JHEP}\
  }\textbf {\bibinfo {volume} {07}},\ \bibinfo {pages} {010} (\bibinfo {year}
  {2020})},\ \Eprint {http://arxiv.org/abs/2004.09536} {arXiv:2004.09536
  [gr-qc]} \BibitemShut {NoStop}%
\bibitem [{\citenamefont {Ernst}(1976)}]{Ernst:1976mzr}%
  \BibitemOpen
  \bibfield  {author} {\bibinfo {author} {\bibfnamefont {F.~J.}\ \bibnamefont
  {Ernst}},\ }\href {\doibase 10.1063/1.522781} {\bibfield  {journal} {\bibinfo
   {journal} {J. Math. Phys.}\ }\textbf {\bibinfo {volume} {17}},\ \bibinfo
  {pages} {54} (\bibinfo {year} {1976})}\BibitemShut {NoStop}%
\bibitem [{\citenamefont {Gibbons}\ \emph {et~al.}(2013)\citenamefont
  {Gibbons}, \citenamefont {Mujtaba},\ and\ \citenamefont
  {Pope}}]{Gibbons:2013yq}%
  \BibitemOpen
  \bibfield  {author} {\bibinfo {author} {\bibfnamefont {G.}~\bibnamefont
  {Gibbons}}, \bibinfo {author} {\bibfnamefont {A.}~\bibnamefont {Mujtaba}}, \
  and\ \bibinfo {author} {\bibfnamefont {C.}~\bibnamefont {Pope}},\ }\href
  {\doibase 10.1088/0264-9381/30/12/125008} {\bibfield  {journal} {\bibinfo
  {journal} {Class. Quant. Grav.}\ }\textbf {\bibinfo {volume} {30}},\ \bibinfo
  {pages} {125008} (\bibinfo {year} {2013})},\ \Eprint
  {http://arxiv.org/abs/1301.3927} {arXiv:1301.3927 [gr-qc]} \BibitemShut
  {NoStop}%
\bibitem [{\citenamefont {Bekenstein}(1973)}]{Bekenstein:1973mi}%
  \BibitemOpen
  \bibfield  {author} {\bibinfo {author} {\bibfnamefont {J.}~\bibnamefont
  {Bekenstein}},\ }\href {\doibase 10.1103/PhysRevD.7.949} {\bibfield
  {journal} {\bibinfo  {journal} {Phys. Rev. D}\ }\textbf {\bibinfo {volume}
  {7}},\ \bibinfo {pages} {949} (\bibinfo {year} {1973})}\BibitemShut {NoStop}%
\bibitem [{\citenamefont {Herdeiro}\ \emph {et~al.}(2013)\citenamefont
  {Herdeiro}, \citenamefont {Degollado},\ and\ \citenamefont
  {R\'unarsson}}]{Herdeiro:2013pia}%
  \BibitemOpen
  \bibfield  {author} {\bibinfo {author} {\bibfnamefont {C.~A.~R.}\
  \bibnamefont {Herdeiro}}, \bibinfo {author} {\bibfnamefont {J.~C.}\
  \bibnamefont {Degollado}}, \ and\ \bibinfo {author} {\bibfnamefont {H.~F.}\
  \bibnamefont {R\'unarsson}},\ }\href {\doibase 10.1103/PhysRevD.88.063003}
  {\bibfield  {journal} {\bibinfo  {journal} {Phys. Rev. D}\ }\textbf {\bibinfo
  {volume} {88}},\ \bibinfo {pages} {063003} (\bibinfo {year} {2013})},\
  \Eprint {http://arxiv.org/abs/1305.5513} {arXiv:1305.5513 [gr-qc]}
  \BibitemShut {NoStop}%
\bibitem [{\citenamefont {Degollado}\ and\ \citenamefont
  {Herdeiro}(2014)}]{Degollado:2013bha}%
  \BibitemOpen
  \bibfield  {author} {\bibinfo {author} {\bibfnamefont {J.~C.}\ \bibnamefont
  {Degollado}}\ and\ \bibinfo {author} {\bibfnamefont {C.~A.}\ \bibnamefont
  {Herdeiro}},\ }\href {\doibase 10.1103/PhysRevD.89.063005} {\bibfield
  {journal} {\bibinfo  {journal} {Phys. Rev. D}\ }\textbf {\bibinfo {volume}
  {89}},\ \bibinfo {pages} {063005} (\bibinfo {year} {2014})},\ \Eprint
  {http://arxiv.org/abs/1312.4579} {arXiv:1312.4579 [gr-qc]} \BibitemShut
  {NoStop}%
\bibitem [{\citenamefont {Sanchis-Gual}\ \emph {et~al.}(2016)\citenamefont
  {Sanchis-Gual}, \citenamefont {Degollado}, \citenamefont {Montero},
  \citenamefont {Font},\ and\ \citenamefont {Herdeiro}}]{Sanchis-Gual:2015lje}%
  \BibitemOpen
  \bibfield  {author} {\bibinfo {author} {\bibfnamefont {N.}~\bibnamefont
  {Sanchis-Gual}}, \bibinfo {author} {\bibfnamefont {J.~C.}\ \bibnamefont
  {Degollado}}, \bibinfo {author} {\bibfnamefont {P.~J.}\ \bibnamefont
  {Montero}}, \bibinfo {author} {\bibfnamefont {J.~A.}\ \bibnamefont {Font}}, \
  and\ \bibinfo {author} {\bibfnamefont {C.}~\bibnamefont {Herdeiro}},\ }\href
  {\doibase 10.1103/PhysRevLett.116.141101} {\bibfield  {journal} {\bibinfo
  {journal} {Phys.\ Rev.\ Lett.}\ }\textbf {\bibinfo {volume} {116}},\ \bibinfo
  {pages} {141101} (\bibinfo {year} {2016})},\ \Eprint
  {http://arxiv.org/abs/1512.05358} {arXiv:1512.05358 [gr-qc]} \BibitemShut
  {NoStop}%
\bibitem [{\citenamefont {Vieira}\ and\ \citenamefont
  {Bezerra}(2016)}]{Vieira:2016ubt}%
  \BibitemOpen
  \bibfield  {author} {\bibinfo {author} {\bibfnamefont {H.}~\bibnamefont
  {Vieira}}\ and\ \bibinfo {author} {\bibfnamefont {V.}~\bibnamefont
  {Bezerra}},\ }\href {\doibase 10.1016/j.aop.2016.06.016} {\bibfield
  {journal} {\bibinfo  {journal} {Annals Phys.}\ }\textbf {\bibinfo {volume}
  {373}},\ \bibinfo {pages} {28} (\bibinfo {year} {2016})},\ \Eprint
  {http://arxiv.org/abs/1603.02233} {arXiv:1603.02233 [gr-qc]} \BibitemShut
  {NoStop}%
\bibitem [{\citenamefont {Hod}(2012)}]{Hod:2012px}%
  \BibitemOpen
  \bibfield  {author} {\bibinfo {author} {\bibfnamefont {S.}~\bibnamefont
  {Hod}},\ }\href {\doibase 10.1103/PhysRevD.86.129902} {\bibfield  {journal}
  {\bibinfo  {journal} {Phys. Rev. D}\ }\textbf {\bibinfo {volume} {86}},\
  \bibinfo {pages} {104026} (\bibinfo {year} {2012})},\ \bibinfo {note}
  {[Erratum: Phys.Rev.D 86, 129902 (2012)]},\ \Eprint
  {http://arxiv.org/abs/1211.3202} {arXiv:1211.3202 [gr-qc]} \BibitemShut
  {NoStop}%
\bibitem [{\citenamefont {Hod}(2013)}]{Hod:2013zza}%
  \BibitemOpen
  \bibfield  {author} {\bibinfo {author} {\bibfnamefont {S.}~\bibnamefont
  {Hod}},\ }\href {\doibase 10.1140/epjc/s10052-013-2378-x} {\bibfield
  {journal} {\bibinfo  {journal} {Eur.\ Phys.\ J.\ C}\ }\textbf {\bibinfo
  {volume} {73}},\ \bibinfo {pages} {2378} (\bibinfo {year} {2013})},\ \Eprint
  {http://arxiv.org/abs/1311.5298} {arXiv:1311.5298 [gr-qc]} \BibitemShut
  {NoStop}%
\bibitem [{\citenamefont {Hod}(2014)}]{Hod:2014baa}%
  \BibitemOpen
  \bibfield  {author} {\bibinfo {author} {\bibfnamefont {S.}~\bibnamefont
  {Hod}},\ }\href {\doibase 10.1103/PhysRevD.90.024051} {\bibfield  {journal}
  {\bibinfo  {journal} {Phys.\ Rev.\ D}\ }\textbf {\bibinfo {volume} {90}},\
  \bibinfo {pages} {024051} (\bibinfo {year} {2014})},\ \Eprint
  {http://arxiv.org/abs/1406.1179} {arXiv:1406.1179 [gr-qc]} \BibitemShut
  {NoStop}%
\bibitem [{\citenamefont {Benone}\ \emph {et~al.}(2014)\citenamefont {Benone},
  \citenamefont {Crispino}, \citenamefont {Herdeiro},\ and\ \citenamefont
  {Radu}}]{Benone:2014ssa}%
  \BibitemOpen
  \bibfield  {author} {\bibinfo {author} {\bibfnamefont {C.~L.}\ \bibnamefont
  {Benone}}, \bibinfo {author} {\bibfnamefont {L.~C.}\ \bibnamefont
  {Crispino}}, \bibinfo {author} {\bibfnamefont {C.}~\bibnamefont {Herdeiro}},
  \ and\ \bibinfo {author} {\bibfnamefont {E.}~\bibnamefont {Radu}},\ }\href
  {\doibase 10.1103/PhysRevD.90.104024} {\bibfield  {journal} {\bibinfo
  {journal} {Phys. Rev. D}\ }\textbf {\bibinfo {volume} {90}},\ \bibinfo
  {pages} {104024} (\bibinfo {year} {2014})},\ \Eprint
  {http://arxiv.org/abs/1409.1593} {arXiv:1409.1593 [gr-qc]} \BibitemShut
  {NoStop}%
\bibitem [{\citenamefont {Wang}\ and\ \citenamefont
  {Herdeiro}(2016)}]{Wang:2015fgp}%
  \BibitemOpen
  \bibfield  {author} {\bibinfo {author} {\bibfnamefont {M.}~\bibnamefont
  {Wang}}\ and\ \bibinfo {author} {\bibfnamefont {C.}~\bibnamefont
  {Herdeiro}},\ }\href {\doibase 10.1103/PhysRevD.93.064066} {\bibfield
  {journal} {\bibinfo  {journal} {Phys. Rev. D}\ }\textbf {\bibinfo {volume}
  {93}},\ \bibinfo {pages} {064066} (\bibinfo {year} {2016})},\ \Eprint
  {http://arxiv.org/abs/1512.02262} {arXiv:1512.02262 [gr-qc]} \BibitemShut
  {NoStop}%
\bibitem [{\citenamefont {Hod}(2015{\natexlab{a}})}]{Hod:2015goa}%
  \BibitemOpen
  \bibfield  {author} {\bibinfo {author} {\bibfnamefont {S.}~\bibnamefont
  {Hod}},\ }\href {\doibase 10.1016/j.physletb.2015.07.071} {\bibfield
  {journal} {\bibinfo  {journal} {Phys.\ Lett.\ B}\ }\textbf {\bibinfo {volume}
  {749}},\ \bibinfo {pages} {167} (\bibinfo {year} {2015}{\natexlab{a}})},\
  \Eprint {http://arxiv.org/abs/1510.05649} {arXiv:1510.05649 [gr-qc]}
  \BibitemShut {NoStop}%
\bibitem [{\citenamefont {Siahaan}(2015)}]{Siahaan:2015xna}%
  \BibitemOpen
  \bibfield  {author} {\bibinfo {author} {\bibfnamefont {H.~M.}\ \bibnamefont
  {Siahaan}},\ }\href {\doibase 10.1142/S0218271815501023} {\bibfield
  {journal} {\bibinfo  {journal} {Int.\ J.\ Mod.\ Phys.\ D}\ }\textbf {\bibinfo
  {volume} {24}},\ \bibinfo {pages} {1550102} (\bibinfo {year} {2015})},\
  \Eprint {http://arxiv.org/abs/1506.03957} {arXiv:1506.03957 [hep-th]}
  \BibitemShut {NoStop}%
\bibitem [{\citenamefont {Hod}(2017)}]{Hod:2016lgi}%
  \BibitemOpen
  \bibfield  {author} {\bibinfo {author} {\bibfnamefont {S.}~\bibnamefont
  {Hod}},\ }\href {\doibase 10.1007/JHEP01(2017)030} {\bibfield  {journal}
  {\bibinfo  {journal} {JHEP}\ }\textbf {\bibinfo {volume} {01}},\ \bibinfo
  {pages} {030} (\bibinfo {year} {2017})},\ \Eprint
  {http://arxiv.org/abs/1612.00014} {arXiv:1612.00014 [hep-th]} \BibitemShut
  {NoStop}%
\bibitem [{\citenamefont {Hod}(2015{\natexlab{b}})}]{Hod:2016yxg}%
  \BibitemOpen
  \bibfield  {author} {\bibinfo {author} {\bibfnamefont {S.}~\bibnamefont
  {Hod}},\ }\href {\doibase 10.1088/0264-9381/32/13/134002} {\bibfield
  {journal} {\bibinfo  {journal} {Class.\ Quant.\ Grav.}\ }\textbf {\bibinfo
  {volume} {32}},\ \bibinfo {pages} {134002} (\bibinfo {year}
  {2015}{\natexlab{b}})},\ \Eprint {http://arxiv.org/abs/1607.00003}
  {arXiv:1607.00003 [gr-qc]} \BibitemShut {NoStop}%
\bibitem [{\citenamefont {Huang}\ and\ \citenamefont
  {Liu}(2016)}]{Huang:2016qnk}%
  \BibitemOpen
  \bibfield  {author} {\bibinfo {author} {\bibfnamefont {Y.}~\bibnamefont
  {Huang}}\ and\ \bibinfo {author} {\bibfnamefont {D.-J.}\ \bibnamefont
  {Liu}},\ }\href {\doibase 10.1103/PhysRevD.94.064030} {\bibfield  {journal}
  {\bibinfo  {journal} {Phys.\ Rev.\ D}\ }\textbf {\bibinfo {volume} {94}},\
  \bibinfo {pages} {064030} (\bibinfo {year} {2016})},\ \Eprint
  {http://arxiv.org/abs/1606.08913} {arXiv:1606.08913 [gr-qc]} \BibitemShut
  {NoStop}%
\bibitem [{\citenamefont {Bernard}(2016)}]{Bernard:2016wqo}%
  \BibitemOpen
  \bibfield  {author} {\bibinfo {author} {\bibfnamefont {C.}~\bibnamefont
  {Bernard}},\ }\href {\doibase 10.1103/PhysRevD.94.085007} {\bibfield
  {journal} {\bibinfo  {journal} {Phys.\ Rev.\ D}\ }\textbf {\bibinfo {volume}
  {94}},\ \bibinfo {pages} {085007} (\bibinfo {year} {2016})},\ \Eprint
  {http://arxiv.org/abs/1608.05974} {arXiv:1608.05974 [gr-qc]} \BibitemShut
  {NoStop}%
\bibitem [{\citenamefont {Sakalli}\ and\ \citenamefont
  {Tokgoz}(2017)}]{Sakalli:2016xoa}%
  \BibitemOpen
  \bibfield  {author} {\bibinfo {author} {\bibfnamefont {I.}~\bibnamefont
  {Sakalli}}\ and\ \bibinfo {author} {\bibfnamefont {G.}~\bibnamefont
  {Tokgoz}},\ }\href {\doibase 10.1088/1361-6382/aa6858} {\bibfield  {journal}
  {\bibinfo  {journal} {Class.\ Quant.\ Grav.}\ }\textbf {\bibinfo {volume}
  {34}},\ \bibinfo {pages} {125007} (\bibinfo {year} {2017})},\ \Eprint
  {http://arxiv.org/abs/1610.09329} {arXiv:1610.09329 [gr-qc]} \BibitemShut
  {NoStop}%
\bibitem [{\citenamefont {Ferreira}\ and\ \citenamefont
  {Herdeiro}(2017)}]{Ferreira:2017cta}%
  \BibitemOpen
  \bibfield  {author} {\bibinfo {author} {\bibfnamefont {H.~R.~C.}\
  \bibnamefont {Ferreira}}\ and\ \bibinfo {author} {\bibfnamefont {C.~A.~R.}\
  \bibnamefont {Herdeiro}},\ }\href {\doibase 10.1016/j.physletb.2017.08.017}
  {\bibfield  {journal} {\bibinfo  {journal} {Phys.\ Lett.\ B}\ }\textbf
  {\bibinfo {volume} {773}},\ \bibinfo {pages} {129} (\bibinfo {year}
  {2017})},\ \Eprint {http://arxiv.org/abs/1707.08133} {arXiv:1707.08133
  [gr-qc]} \BibitemShut {NoStop}%
\bibitem [{\citenamefont {Richartz}\ \emph {et~al.}(2017)\citenamefont
  {Richartz}, \citenamefont {Herdeiro},\ and\ \citenamefont
  {Berti}}]{Richartz:2017qep}%
  \BibitemOpen
  \bibfield  {author} {\bibinfo {author} {\bibfnamefont {M.}~\bibnamefont
  {Richartz}}, \bibinfo {author} {\bibfnamefont {C.~A.~R.}\ \bibnamefont
  {Herdeiro}}, \ and\ \bibinfo {author} {\bibfnamefont {E.}~\bibnamefont
  {Berti}},\ }\href {\doibase 10.1103/PhysRevD.96.044034} {\bibfield  {journal}
  {\bibinfo  {journal} {Phys.\ Rev.\ D}\ }\textbf {\bibinfo {volume} {96}},\
  \bibinfo {pages} {044034} (\bibinfo {year} {2017})},\ \Eprint
  {http://arxiv.org/abs/1706.01112} {arXiv:1706.01112 [gr-qc]} \BibitemShut
  {NoStop}%
\bibitem [{\citenamefont {Huang}\ \emph {et~al.}(2017)\citenamefont {Huang},
  \citenamefont {Liu}, \citenamefont {Zhai},\ and\ \citenamefont
  {Li}}]{Huang:2017whw}%
  \BibitemOpen
  \bibfield  {author} {\bibinfo {author} {\bibfnamefont {Y.}~\bibnamefont
  {Huang}}, \bibinfo {author} {\bibfnamefont {D.-J.}\ \bibnamefont {Liu}},
  \bibinfo {author} {\bibfnamefont {X.-H.}\ \bibnamefont {Zhai}}, \ and\
  \bibinfo {author} {\bibfnamefont {X.-Z.}\ \bibnamefont {Li}},\ }\href
  {\doibase 10.1088/1361-6382/aa7964} {\bibfield  {journal} {\bibinfo
  {journal} {Class.\ Quant.\ Grav.}\ }\textbf {\bibinfo {volume} {34}},\
  \bibinfo {pages} {155002} (\bibinfo {year} {2017})},\ \Eprint
  {http://arxiv.org/abs/1706.04441} {arXiv:1706.04441 [gr-qc]} \BibitemShut
  {NoStop}%
\bibitem [{\citenamefont {Huang}\ \emph {et~al.}(2018)\citenamefont {Huang},
  \citenamefont {Liu}, \citenamefont {Zhai},\ and\ \citenamefont
  {Li}}]{Huang:2018qdl}%
  \BibitemOpen
  \bibfield  {author} {\bibinfo {author} {\bibfnamefont {Y.}~\bibnamefont
  {Huang}}, \bibinfo {author} {\bibfnamefont {D.-J.}\ \bibnamefont {Liu}},
  \bibinfo {author} {\bibfnamefont {X.-h.}\ \bibnamefont {Zhai}}, \ and\
  \bibinfo {author} {\bibfnamefont {X.-z.}\ \bibnamefont {Li}},\ }\href
  {\doibase 10.1103/PhysRevD.98.025021} {\bibfield  {journal} {\bibinfo
  {journal} {Phys.\ Rev.\ D}\ }\textbf {\bibinfo {volume} {98}},\ \bibinfo
  {pages} {025021} (\bibinfo {year} {2018})},\ \Eprint
  {http://arxiv.org/abs/1807.06263} {arXiv:1807.06263 [gr-qc]} \BibitemShut
  {NoStop}%
\bibitem [{\citenamefont {Garc\'{i}a}\ and\ \citenamefont
  {Salgado}(2019)}]{Garcia:2018sjh}%
  \BibitemOpen
  \bibfield  {author} {\bibinfo {author} {\bibfnamefont {G.}~\bibnamefont
  {Garc\'{i}a}}\ and\ \bibinfo {author} {\bibfnamefont {M.}~\bibnamefont
  {Salgado}},\ }\href {\doibase 10.1103/PhysRevD.99.044036} {\bibfield
  {journal} {\bibinfo  {journal} {Phys.\ Rev.\ D}\ }\textbf {\bibinfo {volume}
  {99}},\ \bibinfo {pages} {044036} (\bibinfo {year} {2019})},\ \Eprint
  {http://arxiv.org/abs/1812.05809} {arXiv:1812.05809 [gr-qc]} \BibitemShut
  {NoStop}%
\bibitem [{\citenamefont {Delgado}\ \emph {et~al.}(2019)\citenamefont
  {Delgado}, \citenamefont {Herdeiro},\ and\ \citenamefont
  {Radu}}]{Delgado:2019prc}%
  \BibitemOpen
  \bibfield  {author} {\bibinfo {author} {\bibfnamefont {J.~F.}\ \bibnamefont
  {Delgado}}, \bibinfo {author} {\bibfnamefont {C.~A.}\ \bibnamefont
  {Herdeiro}}, \ and\ \bibinfo {author} {\bibfnamefont {E.}~\bibnamefont
  {Radu}},\ }\href {\doibase 10.1016/j.physletb.2019.04.009} {\bibfield
  {journal} {\bibinfo  {journal} {Phys.\ Lett.\ B}\ }\textbf {\bibinfo {volume}
  {792}},\ \bibinfo {pages} {436} (\bibinfo {year} {2019})},\ \Eprint
  {http://arxiv.org/abs/1903.01488} {arXiv:1903.01488 [gr-qc]} \BibitemShut
  {NoStop}%
\bibitem [{\citenamefont {Kunz}\ \emph {et~al.}(2019)\citenamefont {Kunz},
  \citenamefont {Perapechka},\ and\ \citenamefont {Shnir}}]{Kunz:2019bhm}%
  \BibitemOpen
  \bibfield  {author} {\bibinfo {author} {\bibfnamefont {J.}~\bibnamefont
  {Kunz}}, \bibinfo {author} {\bibfnamefont {I.}~\bibnamefont {Perapechka}}, \
  and\ \bibinfo {author} {\bibfnamefont {Y.}~\bibnamefont {Shnir}},\ }\href
  {\doibase 10.1103/PhysRevD.100.064032} {\bibfield  {journal} {\bibinfo
  {journal} {Phys.\ Rev.\ D}\ }\textbf {\bibinfo {volume} {100}},\ \bibinfo
  {pages} {064032} (\bibinfo {year} {2019})},\ \Eprint
  {http://arxiv.org/abs/1904.07630} {arXiv:1904.07630 [gr-qc]} \BibitemShut
  {NoStop}%
\bibitem [{\citenamefont {Garc\'{i}a}\ and\ \citenamefont
  {Salgado}(2020)}]{Garcia:2019zla}%
  \BibitemOpen
  \bibfield  {author} {\bibinfo {author} {\bibfnamefont {G.}~\bibnamefont
  {Garc\'{i}a}}\ and\ \bibinfo {author} {\bibfnamefont {M.}~\bibnamefont
  {Salgado}},\ }\href {\doibase 10.1103/PhysRevD.101.044040} {\bibfield
  {journal} {\bibinfo  {journal} {Phys.\ Rev.\ D}\ }\textbf {\bibinfo {volume}
  {101}},\ \bibinfo {pages} {044040} (\bibinfo {year} {2020})},\ \Eprint
  {http://arxiv.org/abs/1909.12987} {arXiv:1909.12987 [gr-qc]} \BibitemShut
  {NoStop}%
\bibitem [{\citenamefont {Santos}\ and\ \citenamefont
  {Herdeiro}(2020)}]{Santos:2020sut}%
  \BibitemOpen
  \bibfield  {author} {\bibinfo {author} {\bibfnamefont {N.~M.}\ \bibnamefont
  {Santos}}\ and\ \bibinfo {author} {\bibfnamefont {C.~A.~R.}\ \bibnamefont
  {Herdeiro}},\ }\href {\doibase 10.1142/S0218271820410138} {\bibfield
  {journal} {\bibinfo  {journal} {Int. J. Mod. Phys. D}\ }\textbf {\bibinfo
  {volume} {29}},\ \bibinfo {pages} {2041013} (\bibinfo {year} {2020})},\
  \Eprint {http://arxiv.org/abs/2005.07201} {arXiv:2005.07201 [gr-qc]}
  \BibitemShut {NoStop}%
\bibitem [{\citenamefont {Herdeiro}\ and\ \citenamefont
  {Radu}(2015)}]{Herdeiro:2015gia}%
  \BibitemOpen
  \bibfield  {author} {\bibinfo {author} {\bibfnamefont {C.}~\bibnamefont
  {Herdeiro}}\ and\ \bibinfo {author} {\bibfnamefont {E.}~\bibnamefont
  {Radu}},\ }\href {\doibase 10.1088/0264-9381/32/14/144001} {\bibfield
  {journal} {\bibinfo  {journal} {Class. Quant. Grav.}\ }\textbf {\bibinfo
  {volume} {32}},\ \bibinfo {pages} {144001} (\bibinfo {year} {2015})},\
  \Eprint {http://arxiv.org/abs/1501.04319} {arXiv:1501.04319 [gr-qc]}
  \BibitemShut {NoStop}%
\bibitem [{\citenamefont {Harrison}(1968)}]{Harrison:1971}%
  \BibitemOpen
  \bibfield  {author} {\bibinfo {author} {\bibfnamefont {B.~K.}\ \bibnamefont
  {Harrison}},\ }\href {\doibase 10.1063/1.1664508} {\bibfield  {journal}
  {\bibinfo  {journal} {J. Math. Phys.}\ }\textbf {\bibinfo {volume} {9}},\
  \bibinfo {pages} {1744} (\bibinfo {year} {1968})},\ \Eprint
  {http://arxiv.org/abs/https://doi.org/10.1063/1.1664508}
  {https://doi.org/10.1063/1.1664508} \BibitemShut {NoStop}%
\bibitem [{\citenamefont {{Ernst}}\ and\ \citenamefont
  {{Wild}}(1976)}]{Ernst:1976a}%
  \BibitemOpen
  \bibfield  {author} {\bibinfo {author} {\bibfnamefont {F.~J.}\ \bibnamefont
  {{Ernst}}}\ and\ \bibinfo {author} {\bibfnamefont {W.~J.}\ \bibnamefont
  {{Wild}}},\ }\href {\doibase 10.1063/1.522875} {\bibfield  {journal}
  {\bibinfo  {journal} {J. Math. Phys.}\ }\textbf {\bibinfo {volume} {17}},\
  \bibinfo {pages} {182} (\bibinfo {year} {1976})}\BibitemShut {NoStop}%
\bibitem [{\citenamefont {Konoplya}(2008)}]{Konoplya:2008hj}%
  \BibitemOpen
  \bibfield  {author} {\bibinfo {author} {\bibfnamefont {R.~A.}\ \bibnamefont
  {Konoplya}},\ }\href {\doibase 10.1016/j.physletb.2008.11.059,
  10.1016/j.physletb.2008.07.079} {\bibfield  {journal} {\bibinfo  {journal}
  {Phys. Lett.}\ }\textbf {\bibinfo {volume} {B666}},\ \bibinfo {pages} {283}
  (\bibinfo {year} {2008})},\ \bibinfo {note} {[Phys. Lett.B670,459(2009)]},\
  \Eprint {http://arxiv.org/abs/0801.0846} {arXiv:0801.0846 [hep-th]}
  \BibitemShut {NoStop}%
\bibitem [{\citenamefont {Brito}\ \emph {et~al.}(2014)\citenamefont {Brito},
  \citenamefont {Cardoso},\ and\ \citenamefont {Pani}}]{Brito:2014nja}%
  \BibitemOpen
  \bibfield  {author} {\bibinfo {author} {\bibfnamefont {R.}~\bibnamefont
  {Brito}}, \bibinfo {author} {\bibfnamefont {V.}~\bibnamefont {Cardoso}}, \
  and\ \bibinfo {author} {\bibfnamefont {P.}~\bibnamefont {Pani}},\ }\href
  {\doibase 10.1103/PhysRevD.89.104045} {\bibfield  {journal} {\bibinfo
  {journal} {Phys. Rev. D}\ }\textbf {\bibinfo {volume} {89}},\ \bibinfo
  {pages} {104045} (\bibinfo {year} {2014})},\ \Eprint
  {http://arxiv.org/abs/1405.2098} {arXiv:1405.2098 [gr-qc]} \BibitemShut
  {NoStop}%
\bibitem [{\citenamefont {Cardoso}\ \emph {et~al.}(2008)\citenamefont
  {Cardoso}, \citenamefont {Pani}, \citenamefont {Cadoni},\ and\ \citenamefont
  {Cavaglia}}]{Cardoso:2008kj}%
  \BibitemOpen
  \bibfield  {author} {\bibinfo {author} {\bibfnamefont {V.}~\bibnamefont
  {Cardoso}}, \bibinfo {author} {\bibfnamefont {P.}~\bibnamefont {Pani}},
  \bibinfo {author} {\bibfnamefont {M.}~\bibnamefont {Cadoni}}, \ and\ \bibinfo
  {author} {\bibfnamefont {M.}~\bibnamefont {Cavaglia}},\ }\href {\doibase
  10.1088/0264-9381/25/19/195010} {\bibfield  {journal} {\bibinfo  {journal}
  {Class. Quant. Grav.}\ }\textbf {\bibinfo {volume} {25}},\ \bibinfo {pages}
  {195010} (\bibinfo {year} {2008})},\ \Eprint {http://arxiv.org/abs/0808.1615}
  {arXiv:0808.1615 [gr-qc]} \BibitemShut {NoStop}%
\bibitem [{\citenamefont {Abramowitz}\ and\ \citenamefont
  {Stegun}(1965)}]{mabramowitz64:handbook}%
  \BibitemOpen
  \bibfield  {author} {\bibinfo {author} {\bibfnamefont {M.}~\bibnamefont
  {Abramowitz}}\ and\ \bibinfo {author} {\bibfnamefont {I.~A.}\ \bibnamefont
  {Stegun}},\ }\href@noop {} {\emph {\bibinfo {title} {Handbook of Mathematical
  Functions with Formulas, Graphs and Mathematical Tables}}}\ (\bibinfo
  {publisher} {Dover Publications, Inc.},\ \bibinfo {address} {New York},\
  \bibinfo {year} {1965})\BibitemShut {NoStop}%
\bibitem [{\citenamefont {Pani}(2013)}]{Pani:2013pma}%
  \BibitemOpen
  \bibfield  {author} {\bibinfo {author} {\bibfnamefont {P.}~\bibnamefont
  {Pani}},\ }\href {\doibase 10.1142/S0217751X13400186} {\bibfield  {journal}
  {\bibinfo  {journal} {Int. J. Mod. Phys. A}\ }\textbf {\bibinfo {volume}
  {28}},\ \bibinfo {pages} {1340018} (\bibinfo {year} {2013})},\ \Eprint
  {http://arxiv.org/abs/1305.6759} {arXiv:1305.6759 [gr-qc]} \BibitemShut
  {NoStop}%
\end{thebibliography}%

\end{document}